\documentclass[reprint,aps, nofootinbib, prd, floats, floatfix, amsmath, amssymb, superscriptaddress, preprintnumbers]{revtex4-2}

\usepackage{graphicx}
\usepackage[caption=false]{subfig}
\usepackage{dcolumn}
\usepackage{bm}
\usepackage{amssymb}
\usepackage[utf8]{inputenc}
\usepackage[OT1]{fontenc}
\usepackage{amsmath,amsthm,amsfonts,amssymb,amscd, ulem}
\usepackage{enumerate}
\usepackage{fancyhdr}
\usepackage{mathtools}
\usepackage{mathrsfs}
\usepackage{cancel}
\usepackage{slashed}
\usepackage{bigints}
\usepackage[flushleft]{threeparttable}
\usepackage[colorlinks=true, citecolor=purple, linkcolor=blue]{hyperref}
\usepackage{url}
\usepackage{makecell,booktabs}
\usepackage{braket}
\usepackage{relsize}
\usepackage{multirow}
\usepackage{verbatim}
\usepackage{slashed}
\usepackage{upgreek}
\usepackage{extarrows}
\usepackage{array}
\usepackage{appendix}
\usepackage[T1]{fontenc}
\usepackage{setspace}
\usepackage[dvipsnames]{xcolor}

\usepackage{tikz} 
\usetikzlibrary{shapes,arrows,positioning,automata,backgrounds,calc,er,patterns}
\usepackage{tikz-feynman}
\tikzfeynmanset{compat=1.0.0}
\usetikzlibrary{shapes.misc}
\tikzset{cross/.style={cross out, draw=black, fill=none, minimum size=2*(#1-\pgflinewidth), inner sep=0pt, outer sep=0pt}, cross/.default={2pt}}
\usetikzlibrary{shapes.geometric}
\usepackage{orcidlink}

\renewcommand{\arraystretch}{0.6}

\begin{document}

\title{Ferromagnetic broadband sensing of axionlike dark matter}

 \author{Chenhao Peng}
 \affiliation{School of Physics and State Key Laboratory of Nuclear Physics and Technology, Peking University, Beijing 100871, China}

\author{Dmitry Budker \orcidlink{0000-0002-7356-4814}}
\affiliation{Johannes Gutenberg University, Mainz 55128, Germany}
\affiliation{Helmholtz-Institute, GSI Helmholtzzentrum fur Schwerionenforschung, Mainz 55128, Germany}
\affiliation{Department of Physics, University of California, Berkeley, California 94720, USA}

\author{Yuanning Gao}
\affiliation{School of Physics and State Key Laboratory of Nuclear Physics and Technology, Peking University, Beijing 100871, China}
 \affiliation{Center for High Energy Physics, Peking University, Beijing 100871, China}
\author{Xinran Li}
\affiliation{School of Physics and State Key Laboratory of Nuclear Physics and Technology, Peking University, Beijing 100871, China}
 \affiliation{Center for High Energy Physics, Peking University, Beijing 100871, China}

 \author{Jia Liu \orcidlink{0000-0001-7386-0253}}
 \email[Corresponding author: ]
 {jialiu@pku.edu.cn}
 \affiliation{School of Physics and State Key Laboratory of Nuclear Physics and Technology, Peking University, Beijing 100871, China}
 \affiliation{Center for High Energy Physics, Peking University, Beijing 100871, China}

 \author{Wei Ji \orcidlink{0000-0003-3810-4828}}
	\email[Corresponding author: ]{wei.ji@pku.edu.cn}
	\affiliation{School of Physics and State Key Laboratory of Nuclear Physics and Technology, Peking University, Beijing 100871, China}
    \affiliation{Center for High Energy Physics, Peking University, Beijing 100871, China}

\author{Jing Shu}
\affiliation{School of Physics and State Key Laboratory of Nuclear Physics and Technology, Peking University, Beijing 100871, China}
 \affiliation{Center for High Energy Physics, Peking University, Beijing 100871, China}

 \author{Zhenxing Tang}
\affiliation{School of Physics and State Key Laboratory of Nuclear Physics and Technology, Peking University, Beijing 100871, China}
 \affiliation{Center for High Energy Physics, Peking University, Beijing 100871, China}

 \author{Liheng Wang}
 \affiliation{School of Physics and State Key Laboratory of Nuclear Physics and Technology, Peking University, Beijing 100871, China}

\collaboration{PHOENIX Collaboration}\noaffiliation

\begin{abstract}

Levitated particles have demonstrated ultrahigh sensitivity to magnetic fields and accelerations owing to their extremely low dissipation. 
Such systems have strong potential for fundamental physics research, particularly for the detection of axions and axionlike particles, well-motivated dark matter candidates spanning a broad mass range. In this context, both high sensitivity and large bandwidth are essential. Here, we demonstrate a levitated magnet magnetometer based on an engineered double-resonance mode, achieving an effective linewidth at its optimal sensitivity that is approximately three orders of magnitude broader than those of previous approaches. Together with a hard-magnet array that enhances the axion-induced signal and soft-ferromagnetic shielding that suppresses environmental magnetic noise, this system constitutes a hybrid ferromagnetic platform for axionlike dark matter searches. We search for axionlike dark matter through its photon coupling $g_{a\gamma}$ over the $40\text{–}3000\,\mathrm{Hz}$ frequency range and establish new direct limits in this frequency band.
The best sensitivity is achieved near the upper resonance around $276\,\mathrm{Hz}$, where the magnetometer reaches a magnetic-field resolution of $0.7\,\mathrm{fT}$, corresponding to a limit of $g_{a\gamma}\sim10^{-7}~\mathrm{GeV}^{-1}$. At this frequency, this result improves upon previous direct limits by more than four orders of magnitude. The demonstrated high-bandwidth levitated sensor may also enable a broad range of applications, including biological sensing and precision measurements.
\end{abstract}

\maketitle

Levitated sensors, including both optically and magnetically levitated systems, have demonstrated ultrahigh sensitivity in precision measurements, particularly for force and acceleration sensing, and have become powerful platforms for exploring fundamental quantum phenomena\,\cite{Gonzalez-Ballestero:2021gnu}. Femtotesla-level magnetic sensing using levitated-magnet magnetometer\,(LeMaMa) has only recently been demonstrated in both cryogenic and ambient environments, establishing a new platform for ultralow-dissipation magnetometry\,\cite{Ahrens:2024yzo,Ji:2025yvn}. Owing to their low dissipation and high sensitivity, levitated systems have emerged as promising platforms for precision measurements and fundamental physics searches\,\cite{sheng2026,Amaral:2024rbj,Li:2026rty,Danieli:2026qtx}, particularly for the detection of dark matter candidates such as axions\,\cite{Higgins:2023gwq,Kalia:2024eml}.

Axions and axionlike particles (ALPs) are well-motivated candidates for ultralight dark matter. 
A particularly clear route to their detection is provided by the axion-photon coupling, through which a galactic axion dark-matter field acts as a coherently oscillating classical background and, in the presence of a static magnetic field, induces an oscillating electromagnetic response~\cite{Irastorza:2018dyq,Graham:2013gfa,Sikivie:1983ip,Wilczek:1987mv}. 
This interaction underlies a broad experimental program ranging from particle-conversion searches to direct searches for axion dark matter based on electromagnetic sensing.

Over the past decade, direct searches based on axion electrodynamics have advanced across a wide span of frequencies and sensor technologies. 
Resonant microwave and dielectric haloscopes have reached remarkable sensitivity in the $\mu\mathrm{eV}$ mass range~\cite{ADMX:2024xbv,HAYSTAC:2024jch,MADMAX:2024sxs,DMRadio:2022jfv}, while lumped-element and broadband electromagnetic searches have extended coverage toward lower masses~\cite{Ouellet:2018beu,Salemi:2021gck,DMRadio:2022pkf,Gramolin:2020ict}. 
At the lowest frequencies, terrestrial and geomagnetic-field searches have begun to probe the sub-$40~\mathrm{Hz}$ regime~\cite{Sulai:2023zqw,Friel:2024shg,Nishizawa:2025xka}, leveraging Earth-cavity enhancement. Cryogenic magnetic sensors offer high sensitivity due to large fields and reduced thermal noise, but are often limited at low frequencies by vibrations from refrigeration systems~\cite{Gramolin:2020ict}. Direct laboratory searches remain comparatively sparse in the intermediate $40$--$3000~\mathrm{Hz}$ band, where strong magnetic fields over large volumes, low environmental magnetic noise and sufficiently quiet readout are difficult to achieve simultaneously. 

Recent optical and electromagnetic searches have begun to probe parts of this range~\cite{Oshima:2023csb}, but sensitivity in this intermediate band remains limited.
Near-room-temperature atomic magnetometers have emerged as powerful tools for searches for dark matter and other new physics through axion–spin couplings~\cite{Wei:2023rzs,Xu:2023vfn,Lee:2022vvb,JacksonKimball:2023snl,Cong:2024qly,Budker:2006gya,Dou:2024oxy}. However, their operation becomes challenging in large magnetic fields, limiting their applicability to axion searches based on the axion–photon coupling. By contrast, LeMaMa sensors have recently demonstrated high magnetic sensitivity while operating under Earth's magnetic fields~\cite{Ji:2025yvn}. 
A major challenge for resonant dark-matter searches is that high sensitivity is typically achieved at the expense of bandwidth. The previously demonstrated room-temperature LeMaMa sensor exhibits a resonance linewidth of approximately $0.01\,\mathrm{Hz}$~\cite{Ji:2025yvn}, making it prohibitively time-consuming to scan a broad axion mass range. For searches near $100\,\mathrm{Hz}$, covering the relevant parameter space can require on the order of a year of measurement time, comparable to the timescales associated with resonant scanning in superconducting levitated sensors~\cite{Kalia:2024eml,Higgins:2023gwq}. Here, we overcome this limitation by engineering two librational modes and tailoring both their dissipation and resonance frequencies. The resulting double-resonance system exhibits an effective linewidth of approximately $12\,\mathrm{Hz}$, extending the accessible bandwidth by more than three orders of magnitude while maintaining high magnetic sensitivity.

Ferromagnetic materials provide a versatile platform whose functionality can be engineered through magnetic properties such as coercivity. Here, we demonstrate a hybrid ferromagnetic architecture that simultaneously enhances axion-induced signals and suppresses environmental magnetic noise. The system combines a hard-ferromagnet ring array that generates the static magnetic field required for axion-induced effective currents, a soft-ferromagnetic shield that attenuates ambient magnetic noise, and a levitated ferromagnetic sensor that provides ultrasensitive readout. Operating entirely at room temperature avoids vibration noise associated with cryogenic refrigeration, while the use of insulating permanent magnets eliminates magnetic noise arising from current fluctuations.

With this approach, we perform a direct search for axion dark matter through the axion--photon coupling $g_{a\gamma}$ in the $40$--$3000\,\mathrm{Hz}$ frequency range and set new laboratory bounds. 
Near the most sensitive region around $276\,\mathrm{Hz}$, we reach a magnetic resolution of $0.7\,\rm{fT}$ and a coupling sensitivity of $g_{a\gamma}\sim 10^{-7}\,\mathrm{GeV}^{-1}$, which represents an improvement of more than four orders of magnitude over previous direct searches for axionlike particle dark matter, with a potential for future improvement beyond the level of indirect limits. Beyond axion searches, the same broadband LeMaMa architecture can be readily adapted to a wide range of sensing applications, including biological magnetic-signal detection and searches for new physics, such as exotic spin-dependent interactions.
\\

\noindent \textbf{\large Experimental setup and detector characterization}\\

\newlength{\figH}
\setlength{\figH}{0.32\textheight}
\newlength{\leftW}
\newlength{\rightW}
\setlength{\leftW}{0.54\textwidth}
\setlength{\rightW}{0.46\textwidth}
\begin{figure*}[tpbh]
    \centering
    \begin{minipage}[t]{\leftW}
        \centering
        \subfloat[]{%
            \begin{minipage}[b][\figH][c]{\linewidth}
                \makebox[\linewidth][r]{%
                    \includegraphics[
                        width=\linewidth,
                        height=\figH,
                        keepaspectratio,
                        trim=2cm 0cm 2cm 0cm,
                        clip
                    ]{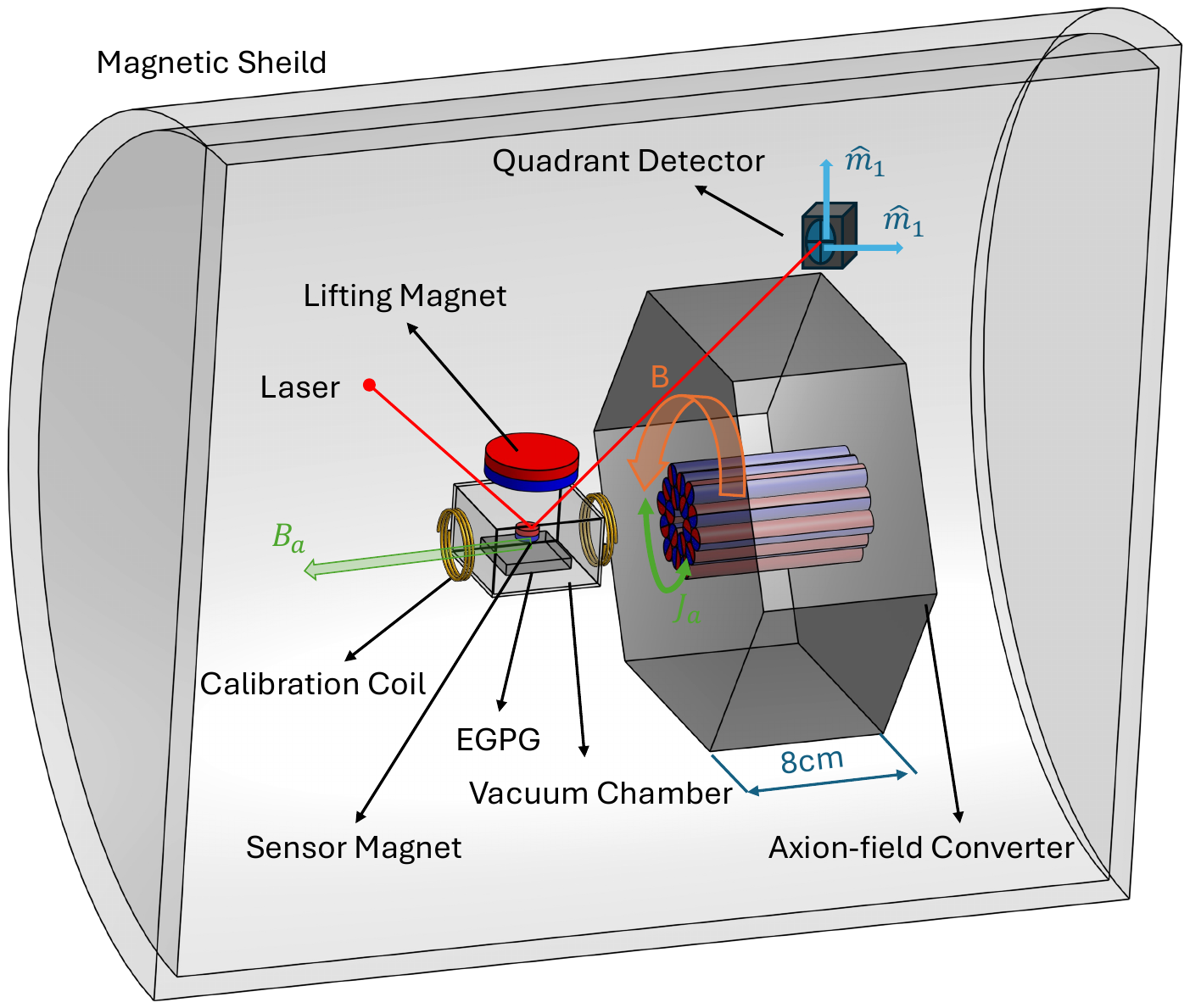}%
                }%
            \end{minipage}%
            \label{fig:source_a}%
        }%
    \end{minipage}%
    \hspace{0pt}%
    \begin{minipage}[t]{\rightW}
        \centering
        \subfloat[]{%
            \begin{minipage}[b][\figH][c]{\linewidth}
                \makebox[\linewidth][l]{%
                    \includegraphics[
                        width=\linewidth,
                        height=\figH,
                        keepaspectratio,
                        trim=3.5cm 0cm 0.5cm 3cm,
                        clip
                    ]{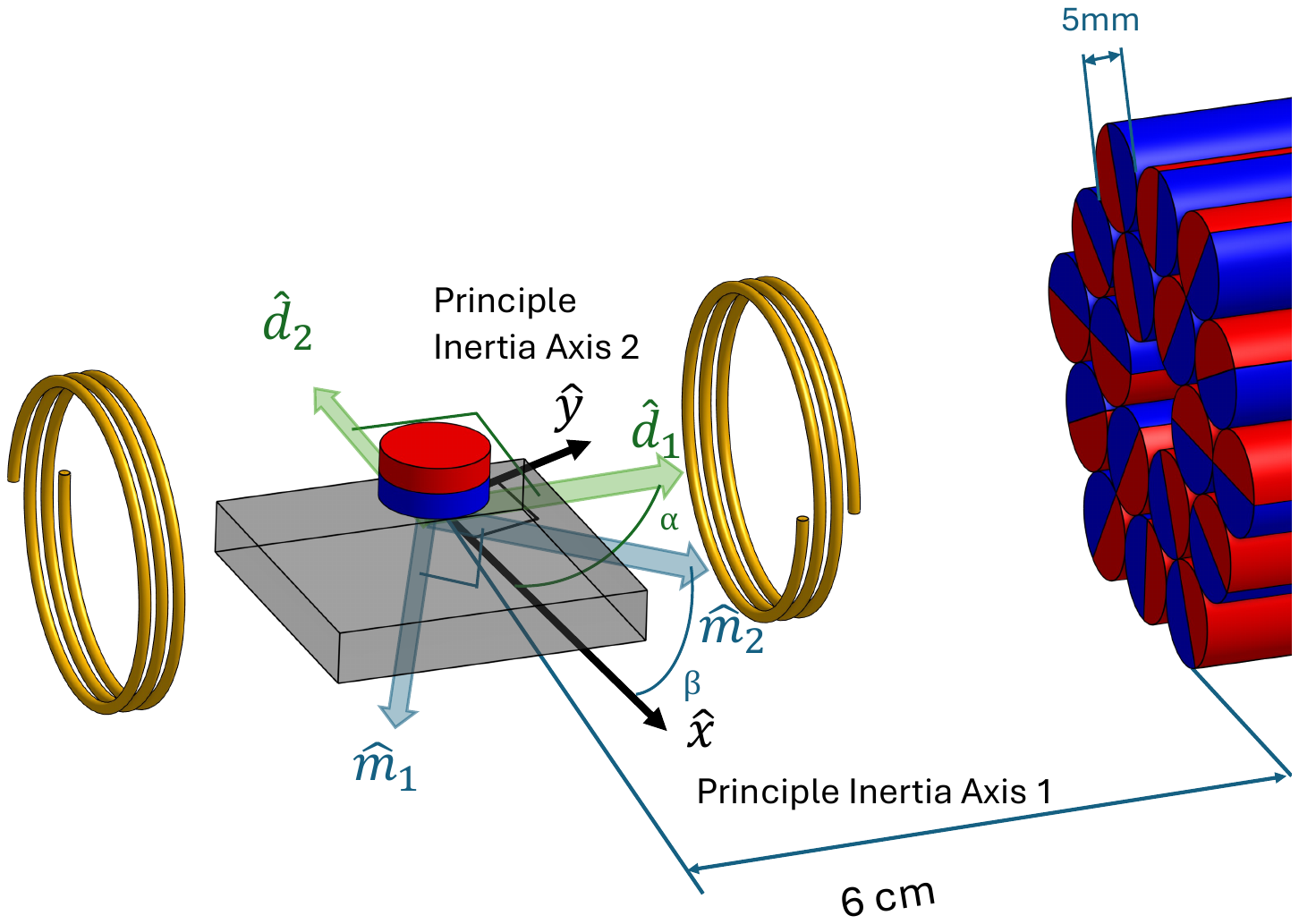}%
                }%
            \end{minipage}%
            \label{fig:source_b}%
        }%
    \end{minipage}
    \vspace{10pt}
    \caption{Schematic of the experimental setup, the geometry of the two-mode response model, and the converter-magnet arrangement. (a) Experimental setup, including the toroidally arranged ferrite permanent-magnet converter array, magnetic shielding, and the levitated ferromagnetic magnetometer (LeMaMa) readout. The converter array is assembled from 13,248 cylindrical ferrite magnets arranged in a hexagonal close-packed structure around a central bore, forming 23 concentric hexagonal rings and eight stacked layers; for visual clarity, only the two innermost rings are shown. The whole setup is mounted on a vibration-isolation stage to reduce seismic noise. (b) Geometry of the two-mode response model in a fixed modal basis. The injected calibration drive is applied along $\hat{\mathbf d}_1 $, and its orthogonal direction $\hat{\mathbf d}_2$ is used to parameterize the noise entering transverse to the signal direction. The two readout channels correspond to orthonormal detection directions $\hat{\mathbf m}_1$ and $\hat{\mathbf m}_2$.}
    \label{fig:source}
\end{figure*}

A schematic of the experimental setup is shown in Fig.~\ref{fig:source}. The LeMaMa sensor magnet is cylindrical, with dimensions of $r \sim 500\,\mu\mathrm{m}$ and $h \sim 200\,\mu\mathrm{m}$. It is levitated by a lifting magnet positioned above an epoxy-glued pyrolytic graphite (EGPG) layer below, which provides diamagnetic stabilization\,\cite{Ji:2025yvn}. The mechanical motion is read out by reflecting a $780~\mathrm{nm}$ probe laser from the top surface of the levitated magnet onto a quadrant photo detector (QPD), which provides two simultaneous readout channels (along $\hat{m}_1$ and $\hat{m}_2$) and reconstructs the two-dimensional angular motion of the sensor. The levitated sensor and EGPG are placed inside a vacuum chamber to suppress air-convection noise.

A central procedure in the experiment is the direct calibration of the magnetic-field response. To this end, we employ calibration coils surrounding the sensor region to generate a controlled oscillating magnetic field with programmable amplitude and frequency. The calibration field is aligned coaxially with the initial direction of the laser beam, the magnetic shielding, and the axion-field converter magnet. The expected axion-induced signal ($\vec{B}_a$ in Fig.\,\ref{fig:source}) is oriented along the same direction, denoted by $\hat{\mathbf d}_1$ in Fig.\,\ref{fig:source}. The magnet response is measured simultaneously in both QPD channels.

In the presence of a horizontal oscillating magnetic field, the levitated magnet undergoes two in-plane rotational eigenmodes, which define the fixed modal basis denoted by $\hat{x}$ and $\hat{y}$ in Fig.~\ref{fig:source}.  Because of the slight asymmetry and misalignment, a drive applied along the signal direction $\hat{d}_1$ generally excites both modes, and each readout channel of QPD at $\hat{m}_1$ and $\hat{m}_2$ receives a different linear combination of the two. We therefore model the detector as a two-mode linear system in a fixed modal basis. The drive basis is defined by the signal direction $\hat{\mathbf d}_1=(\cos\alpha,\sin\alpha)$ and its orthogonal complement $\hat{\mathbf d}_2=(-\sin\alpha,\cos\alpha)$, while the two QPD readout channels define the orthonormal detection directions $\hat{\mathbf m}_1=(\sin\beta,-\cos\beta)$ and $\hat{\mathbf m}_2=(\cos\beta,\sin\beta)$, as shown in Fig.\,\ref{fig:source}.

Each mode is described by a complex Lorentzian transfer function, $F_i(\omega) = Q_i/\big((\omega_i^2-\omega^2)Q_i-i\,\omega_i\,\omega\big)$, with $Q_i$ being the quality factor for the two modes. The full detector response is encoded in a $2\times 2$ transfer matrix,
\begin{align}
H_{ij} = k_i\,\hat{\mathbf m}_i \cdot
\begin{pmatrix}
F_1(\omega)&0\\
0&F_2(\omega)
\end{pmatrix} \cdot
\hat{\mathbf d}_j^T\,,
\label{eq:fit-model}
\end{align}
where $j$ labels the input direction, $i$ labels the readout channel and $k_i$ represents the efficiency for each readout channel. Equation \eqref{eq:fit-model} defines our response model, which contains eight parameters: two resonance frequencies $f_{1,2}\equiv \omega_{1,2}/(2\pi)$, two readout efficiencies $k_{1,2}$, two quality factors $Q_{1,2}$, the signal-direction angle $\alpha$, and the detection-direction angle $\beta$.

\begin{figure}[tpb]
    \centering
    \includegraphics[width=1\linewidth]{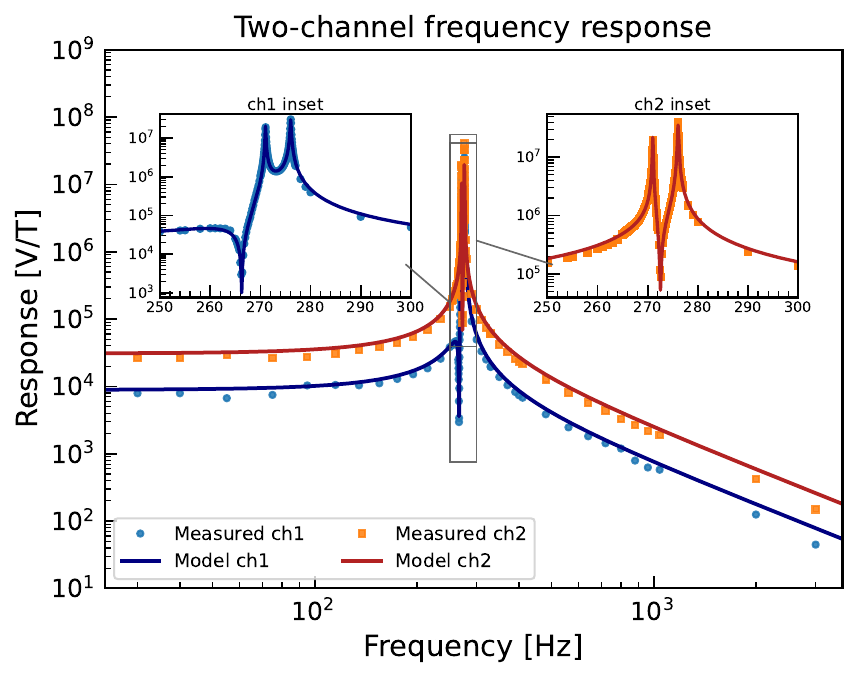}
    \caption{Measured responses $|H_{11}(\omega)|$ (ch1) and $|H_{21}(\omega)|$ (ch2) for a calibration drive applied along $\hat{\mathbf d}_1$, together with the best-fit two-mode model. Points denote the measured data and solid lines denote the eight-parameter model fit. The insets show zoomed-in views of the resonance region, where two nearby resonances associated with the split in-plane rotational modes of the levitated sensor are clearly resolved. The narrow dips arise from destructive interference between the two-mode contributions after projection onto the corresponding readout channel.}
    \label{fig:response}
\end{figure}

In the calibration run, we inject the drive field along $\hat{\mathbf d}_1$ and fit the measured responses in both readout channels, $|H_{11}(\omega)|$ and $|H_{21}(\omega)|$. 
The measured responses and best-fit curves are shown in Fig.\,\ref{fig:response}.
Because the levitated cylindrical magnet is not perfectly symmetric, the two orthogonal  rotational eigenmodes are slightly split in frequency, shown as two distinct resonance peaks. From the best-fit response model, we extract the resonance frequencies
$f_1 = 271.02(2)\,\mathrm{Hz}$ and 
$f_2 = 276.04(2)\,\mathrm{Hz}$.
Other response-model parameters are summarized in Table~\ref{tab:response_params}. Utilizing the two resonance peaks enables broadband detection. 

\begin{table}[tpb]
\centering
\caption{Best-fit response-model parameters obtained from a joint fit to the measured magnitude responses $|H_{11}(\omega)|$ and $|H_{21}(\omega)|$ (Figs.~\ref{fig:response}). For the two modes, we report the resonant frequencies $f_i\equiv\omega_i/(2\pi)$, quality factors $Q_i$, overall channel gains $k_i$, and the geometric parameters $(\alpha,\beta)$ controlling the projection between the drive/readout bases (Fig.~\ref{fig:source}).}
\label{tab:response_params}
\renewcommand{\arraystretch}{1.15}
\begin{tabular}{l | c}
\hline\hline
Parameter & Best-fit value \\
\hline
$f_1$ [Hz] & $271.02(2)$ \\
$f_2$ [Hz] & $276.04(2)$ \\
$Q_1$ & $2.3(2)\times 10^{3}$ \\
$Q_2$ & $1.7(2)\times 10^{3}$ \\
$k_1$ [$\mathrm{V}\cdot \mathrm{Hz}^2/\mathrm{T}$] & $8.5(8)\times 10^{10}$ \\
$k_2$ [$\mathrm{V}\cdot \mathrm{Hz}^2/\mathrm{T}$] & $9.7(5)\times 10^{10}$ \\
$\tan\alpha$ & $2.2(2)$ \\
$\tan\beta$ & $1.06(2)$ \\
$\alpha$ [deg] & $66(2)$ \\
$\beta$ [deg] & $46.6(6)$ \\
\hline\hline
\end{tabular}
\end{table}

The zoomed view in Fig.\,\ref{fig:response} reveals a narrow suppression point close to the resonances, which arises from destructive interference between the two mode contributions after projection onto a single readout axis. Capturing this interference is important for an accurate description of the frequency-dependent signal response. 
More generally, the good agreement between the measured calibration data and the fitted two-mode model demonstrates that the detector response is well understood. Details of the calibration procedure are provided in the Methods.

The choice of parameters for dark matter searches is guided by the resonance frequency and linewidth. The resonance frequency is inversely proportional to the size of the sensor. When the levitated sensor magnet is large, i.e. the resonance frequency is low, low-frequency vibration noise becomes more prominent and degrades the sensitivity. On the other hand, if the sensor magnet is too small, it suffers from degraded optical readout performance arising from diffraction limits and diminished reflected-light intensity. For the best performance we choose a radius of $\sim 500~\mu\mathrm{m}$, corresponding to resonance frequencies around $276~\mathrm{Hz}$. The linewidth is tuned by changing the $Q$ value. A higher quality factor $Q$ yields a greater resonant response and generally improves peak sensitivity. However, for broadband operation—as adopted in this work—a larger $Q$ conversely reduces the intrinsic magnetic response linewidth. In particular, when ambient noise couples to the sensor, elevated $Q$ further amplifies such unwanted noise. A more effective strategy to balance resonant sensitivity and bandwidth is to broaden the linewidth via moderate $Q$ reduction. Additionally, high-$Q$ devices suffer much faster sensitivity degradation above the resonance frequency, owing to the steeper roll-off of the resonant response curve. 
Related discussions of linewidth and sensitivity broadening in high-$Q$ resonant searches can be found in Refs.~\cite{Chaudhuri:2018rqn, Ji:2025yvn, Xu:2023vfn, Aybas:2026rwu}.

\begin{figure}[htb]
    \centering
    \includegraphics[width=0.9\linewidth]{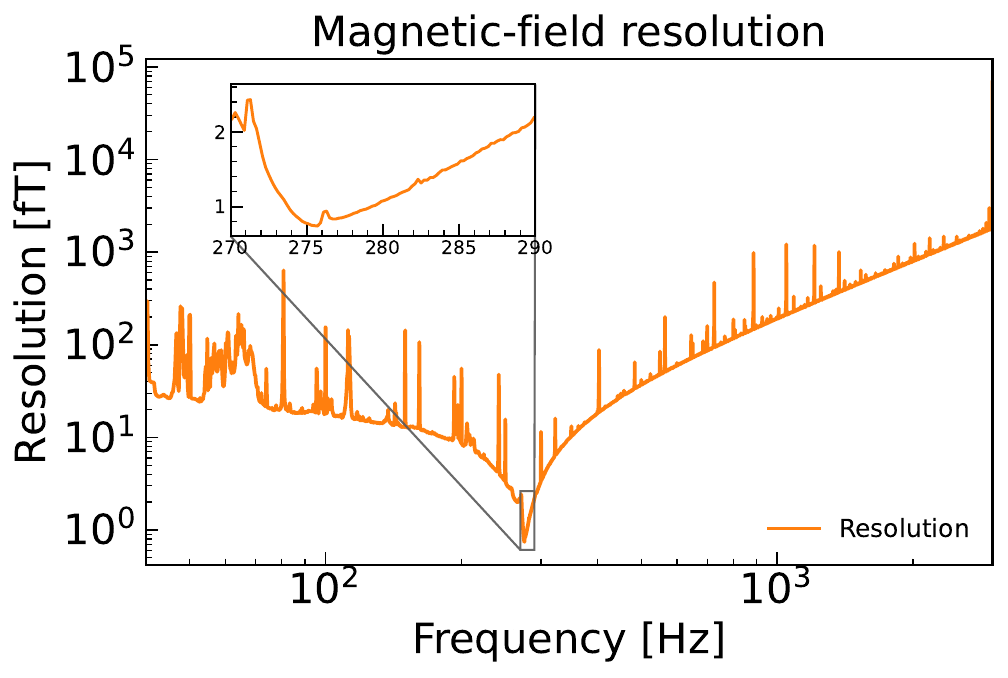}
    \caption{Magnetic-field resolution $R(\omega)$ inferred from the fitted response $\eta(\omega)$ and the measured noise spectrum, with 0.2-Hz bin averaging. The non-smooth structures in the curve mainly reflect excess technical noise, including low-frequency vibrational modes of the supporting frame, 50\,Hz power-line harmonics, and 80\,Hz harmonics likely associated with the data-acquisition chassis.}
    \label{fig:resolution}
\end{figure}

To quantify the effective sensitivity linewidth enabled by the double-resonance modes, it is first necessary to characterize the detector noise. In addition to calibration runs with injected magnetic-field signals, we perform measurements without any signal injection to determine the intrinsic noise level of the system. These measurements also reproduce the operating conditions of the dark-matter search, in which no externally injected signals are present. The resulting noise spectrum therefore provides a direct assessment of the sensitivity and bandwidth achievable in the search. The noise is described by two components: an uncoupled white noise ($\sigma_{\rm unc}$) that enters independently in each readout channel, and a coupled white noise at the input ($\sigma_{\rm c}$)
from the orthogonal directions $(\hat{\mathbf d}_1,\hat{\mathbf d}_2)$. The latter noise is shaped by the transfer functions, whereas the former sets a broadband noise floor. Their relative contributions determine the effective sensitivity linewidth~\cite{Xu:2023vfn}. This simple model captures the main resonant enhancement of the noise as well as the broadband baseline in the measured spectra.

With the fitted response model and noise model, we obtain the magnetic resolution spectrum from 15\,h of measurement data, as shown in Fig.\,\ref{fig:resolution}; see the response and noise section in Methods for details of the fitting procedure.
An effective sensitivity linewidth of $\Gamma_{\mathrm{FWHM}}^{\mathrm{sen}}\simeq 2\pi\cdot12\,\mathrm{Hz}$ is obtained near the most sensitive region. This effective sensitivity linewidth is determined by the combined response of the two resonance modes and the relative contributions of the resonantly shaped and broadband noise components.
This corresponds to an enhancement of nearly three orders of magnitude in bandwidth compared with previously demonstrated LeMaMa sensors that rely on single-mode and high-quality-factor resonances\,\cite{Ji:2025yvn}. Our best magnetic-field resolution is approximately $0.7\,\mathrm{fT}$ near the upper resonance around $276\,\mathrm{Hz}$. 
Away from resonance, the detector retains broadband sensitivity, with a magnetic-field resolution of approximately $20\,\mathrm{fT}$ over approximately $130$--$250~\mathrm{Hz}$.
The stability of the system was verified through repeated response sweeps over a 24\,h period. These measurements show that the calibrated resonance structure remains stable over the integration time relevant to the search. Further details are provided in the Response stability section in Methods. The calibrated response parameters and noise inputs are carried directly into the axion signal analysis.
\\

\noindent \textbf{\large Dark matter signal model and analysis procedure}\\

Ultralight axionlike particles produced non-thermally in the early Universe can be described as a classical oscillating field. Locally, one may write the axion field in a volume $V$ as\,\cite{Lee:2022vvb}
\begin{equation}
a(x)=\sum_{\bf {p}}\sqrt{\frac{2N_{\bf p}}{V \omega_{\bf p}}}\cos(\omega_{\bf p}t-{\bf p}\cdot \bf{x}+\phi_{\bf p})\, 
\label{eq:axionfield}
\end{equation}
with $\omega_{\bf p}\simeq m_a$, where $m_a$ is the ALP mass and $\phi_{\bf p}$ is an a priori unknown phase for momentum mode $\bf p$. $N_{\bf p}$ is the occupation number for mode ${\bf p}$ and is proportional to the DM velocity distribution function. Over times short compared with the coherence time, the field may be treated as approximately monochromatic as\,\cite{Gramolin:2021mqv} 
\begin{equation}
a(t,\mathbf{x}) \simeq \bar a \cos(m_a t)\,,
\end{equation}
where
$\bar a = \sqrt{2\rho_{\rm DM}}/m_a$ and $\rho_{\rm DM}\simeq 0.4~\mathrm{GeV}\,\mathrm{cm}^{-3}$ is the local dark-matter density~\cite{Catena:2009mf}. In the Standard Halo Model, the finite virial velocity dispersion broadens the signal into a narrow spectral line with fractional width $\Delta\omega/\omega\sim10^{-6}$.

The axion--photon interaction,
$\mathcal{L}\supset \frac{1}{4}g_{a\gamma}a F_{\mu\nu}\widetilde{F}^{\mu\nu},$
modifies Maxwell's equations such that, in the presence of a static magnetic field $\mathbf{B}_0$, the axion field acts as an effective oscillating current density~\cite{Kahn:2016aff,Sikivie:2013laa} 
\begin{equation}
\label{Eq:EffCurr}
\mathbf{J}_{\mathrm{a}}=-g_{a\gamma}\dot a\,\mathbf{B}_0\,.
\end{equation}

In this work, a set of hard magnets is used to generate a predominantly toroidal static field, and thus, an effective current according to Eq.\,\eqref{Eq:EffCurr}. The toroidally arranged ferrite permanent-magnet array is shown schematically in Fig.\,\ref{fig:source}.
This axion-driven current produces an oscillating magnetic field $B_a$ at the sensor location, which is then detected with the levitated sensor. 

The array is constructed from 13,248 cylindrical ferrite magnets, each with radius $r_0=5~\mathrm{mm}$ and length $l_0=1\,\mathrm{cm}$. Each magnet is magnetized transversely to the axis of the cylinder. The magnets are arranged in a hexagonal close-packed structure around a central bore, with the innermost ring containing six magnets and the second ring containing twelve. The full transverse structure extends over 23 concentric hexagonal rings. Eight such layers are then stacked to form the complete converter. The ferrite magnets have remanence $B_{\rm r}\approx0.45\,\mathrm{T}$. After assembly, the magnetic flux is approximately confined in closed loops around the bore, producing a well-defined toroidal field configuration while strongly suppressing leakage outside the converter. The separation between the converter and the levitated sensor is $6\,\mathrm{cm}$, and at the sensor position, the residual stray field from the converter is below $0.5\,\mathrm{mT}$.

Ferrite magnets are used instead of neodymium to reduce thermal current noise because of the higher resistivity, while the support structures are made of plastic for the same reason. For the employed experimental geometry, numerical simulation gives an effective source field of order $\mathbf{B}_0\sim \mathcal{O}(0.26\,\mathrm{T})$ over the relevant conversion volume. The corresponding axion-induced signal field at the sensor is nominally horizontal, which defines the signal-field direction used in the detector calibration described below. 

We parameterize the axion-induced magnetic field as
\begin{equation}\label{eq:convertMagneticField}
B_a=\mathcal{K}\,m_a\,g_{a\gamma} \bar{a}
=2.47\times 10^{-8}\, g_{a\gamma}~\mathrm{T\cdot GeV},
\end{equation}
where $\mathcal{K}$ is a geometry factor  for the setup, 
\begin{align}\label{eq:Ba_omega}
\mathcal K
\equiv
\frac{\mu_0}{4\pi}
\left|
\hat{\mathbf d}_1\cdot
\int_{} d^3\mathbf r\,
\mathbf B_0(\mathbf r)\times
\frac{\mathbf r_s-\mathbf r}{|\mathbf r_s-\mathbf r|^3}
\right|,
\end{align}
where $\mathbf r_s$ is the sensor location, and $\mathbf B_0(\mathbf r)$ is the converter magnetic field. $\mathcal{K}$ is numerically calculated using COMSOL Multiphysics~\cite{comsol64}
and treated as a fixed input in the analysis.

Following Eq.\,\eqref{eq:axionfield}, we model galactic axion dark matter as a stochastic classical field with a Maxwell--Boltzmann velocity distribution in the Standard Halo Model. In the large-occupation limit, the Fourier components of the signal are well approximated as zero-mean Gaussian random variables with a narrow, velocity-broadened covariance profile. In the frequency-domain likelihood analysis, this signal covariance is combined with the calibrated detector response. 
The resulting signal covariance matrix takes the form
\begin{equation}
\hat{\bm{\Sigma}}_a(\omega;g_{a\gamma},m_a)=\Sigma_a(\omega;g_{a\gamma},m_a)\,\eta^2(\omega)\,,
\end{equation}
where $\Sigma_a$ is the intrinsic axion spectral variance in magnetic-field units. For the dark-matter search, the two QPD readout channels are combined from the orthogonal channels, reflecting the signal drive from the $\hat{\mathbf d}_1$ direction. Therefore, $\eta(\omega) \equiv \sqrt{|H_{11}(\omega)|^2+|H_{21}(\omega)|^2}$ is the calibrated detector response for the combined channel. The explicit expression for $\Sigma_a$ and the derivation of the stochastic signal covariance are given in the Response and noise section in Methods.

For each scanned axion mass point, we work with the real FFT quadratures of the combined readout and model them as a zero-mean multivariate Gaussian with covariance equal to the sum of the signal and background contributions. The background covariance is constructed from the calibrated noise model. We then use a profile-likelihood-ratio analysis to test the background-only hypothesis and to set upper limits on the non-negative signal-strength parameter $\mu\equiv g_{a\gamma}^2$~\cite{Cowan:2010js}. In the absence of a significant excess, we derive one-sided $95\%$ confidence-level upper limits on $g_{a\gamma}$ at each scanned mass point.

We scan over axion frequencies on a grid with spacing $\Delta f_a/2$~\cite{Lee:2022vvb}, where $\Delta f_a \equiv f_a v_0^2$ denotes the characteristic linewidth arising from the kinetic-energy distribution of virialized axion dark matter in the Standard Halo Model. Here $v_0\simeq 220~\mathrm{km/s}$ sets the characteristic velocity scale of the halo dark-matter distribution.
For each tested frequency $f_a$, the likelihood is evaluated in a finite asymmetric frequency window around $f_a$. 
This overlapping scan grid produces a large set of correlated axion mass hypotheses. Because we test so many distinct hypotheses across the full band, the discovery threshold must be corrected for the look-elsewhere effect. Since the hypotheses are not independent, we calibrate the effective number of independent trials using null Monte Carlo simulations generated with the measured response and noise model, and use the resulting global threshold to assess discovery significance across the full scan band. The details of the profile-likelihood construction, asymptotic test statistics, look-elsewhere simulation, and limit-setting procedure are summarized in Supplementary Note 1.

We further validate the search pipeline by injecting simulated axionlike signals into the measured time series and re-running the likelihood analysis. The recovered couplings are consistent with the injected values. This confirms that the statistical framework can correctly identify and reconstruct weak narrowband axionlike signals. A representative injection-and-recovery example is shown in Supplementary Fig.\,S4.
\\

\noindent \textbf{\large Results and Discussions}\\
The dark matter search experiment accumulated approximately 150 hours of data, which were divided into six datasets. Datasets 1 and 2 were used for detector calibration and system-stability characterization. Dataset 3 was reserved for the final axion-model test and limit setting, and was also used for the sensitivity analysis shown in Fig.\,\ref{fig:resolution}. Datasets 4 , 5 and 6 served as independent validation samples for identifying spurious-signal veto criteria.

Dataset~3 was acquired in a nominal 15\,h run. After vetoing a 10\,min interval affected by abnormal fluctuations, the effective live time was 14\,h~50\,min. In the scan of Dataset~3, we identified 106 narrowband candidate outliers with global significance exceeding $5\sigma$ according to the profile-likelihood-ratio test. To test whether these candidates were associated with persistent instrumental features, rather than statistical fluctuations or axionlike signals, we applied the same analysis procedure to the independent validation datasets, Dataset 4--6. We found that these candidates are not persistent axionlike signals, but are instead more likely attributable to run-dependent instrumental backgrounds, as they did not persist across later experiments acquired at different times. We therefore excluded these candidates from the analysis. Details of their spectrum are illustrated in Supplementary Fig.~S5.

\begin{figure}[tpb]
    \centering
    \includegraphics[width=1\linewidth]{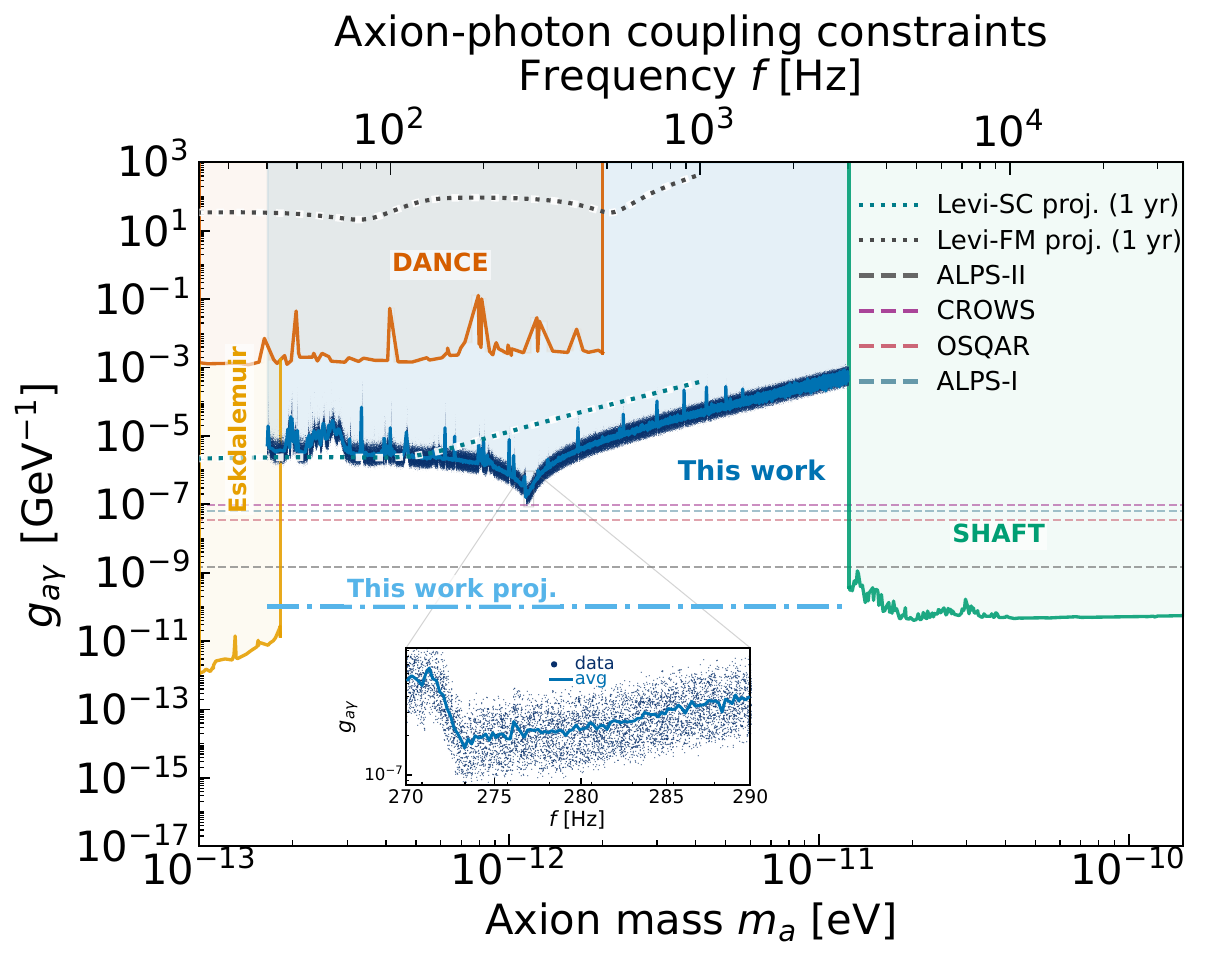}
    \caption{Constraints on the axion--photon coupling $g_{a\gamma}$ obtained with LeMaMa. Blue points show the derived $95\%$ C.L.\ upper limits in the frequency range $40$--$3000~\mathrm{Hz}$.  The solid-blue curve is a $0.2~\mathrm{Hz}$ moving average to better illustrate the overall trend of the exclusion curve. We use solid lines to denote results from direct dark-matter searches~\cite{Gramolin:2020ict,Nishizawa:2025xka,Oshima:2023csb}, and dashed lines to denote non-dark-matter constraints~\cite{Betz:2013dza,Ballou:2014myz,Ehret:2010mh,ALPSII:2025eri}. The dash-dotted blue line denotes the projection for the present experiment, while the two additional proposal curves show the one-year projected sensitivities from previous levitated-sensor proposals~\cite{Higgins:2023gwq,Kalia:2024eml}, labeled as ``Levi-SC proj. (1 yr)'' and ``Levi-FM proj. (1 yr)''.}
    \label{fig:constraint}
\end{figure}

We therefore find no evidence for an axion signal and present the resulting $95\%$ C.L.\ upper limits over the frequency range $40$--$3000\,\mathrm{Hz}$, as shown in Fig.~\ref{fig:constraint}.
As expected, the sensitivity is strongest in the resonance region near the fitted mode frequencies $f_1$ and $f_2$ (Table~\ref{tab:response_params}), where the calibrated response is maximal, and gradually weakens away from resonance. 
Using the 0.2-Hz-averaged exclusion curve, the strongest limit is reached near
$f \simeq 273.45~\mathrm{Hz}$,
\begin{equation}
g_{a\gamma} < 1.8\times 10^{-7}~\mathrm{GeV}^{-1}\,, \text{(95\% C.L.)}.
\end{equation}
Moreover, the limit remains below $7.0\times 10^{-7}~\mathrm{GeV}^{-1}$ over the frequency interval $270$--$282~\mathrm{Hz}$, corresponding to a magnetic-field resolution better than $2.7\,\mathrm{fT}$.

As a direct axion dark-matter search in the same low-frequency range, our limits improve upon the dark-matter-motivated constraint from DANCE~\cite{Oshima:2023csb} by more than four orders of magnitude. To the best of our knowledge, they also provide the first direct laboratory axion-dark-matter limits on $g_{a\gamma}$ in the frequency range from $500\,\mathrm{Hz}$ to $3000\,\mathrm{Hz}$~\cite{AxionLimits}.
We also compare our results with projected sensitivities from previous levitated-sensor proposals based on experimentally demonstrated detector sensitivities~\cite{Higgins:2023gwq,Kalia:2024eml}.
In Fig.~\ref{fig:constraint}, these are labeled as ``Levi-SC proj. (1 yr)'' for the levitated-superconductor proposal and ``Levi-FM proj. (1 yr)'' for the proposal based on superconducting levitation of a ferromagnet, both assuming an integration time of one year with their broadband mode.
Above $200\,\mathrm{Hz}$, our current result already surpasses the levitated-superconductor projection by more than one order of magnitude and exceeds the superconducting-levitation ferromagnet projection by over six orders of magnitude.
This comparison is especially notable because the present analysis is based on only 15\,h of data.
These results highlight the strong potential of the present room-temperature hybrid ferromagnetic platform for future dark-matter searches.

A key advantage of the LeMaMa-based approach is its clear and scalable upgrade path. First, LeMaMa is an emerging technique, whose magnetic field sensitivity has only recently been demonstrated\,\cite{Ji:2025yvn,Ahrens:2024yzo}. Substantial improvements are expected under optimized conditions, such as higher vacuum to suppress Brownian noise and enhanced optical reflectivity to reduce shot noise\,\cite{Ji:2025yvn}.
Second, the axion-induced field at the sensor scales with both the converter magnetic field and the geometric coupling between the converter and the sensor. Enlarging the permanent-magnet converter array and reducing the converter--sensor separation in a more compact design could provide an additional $\mathcal{O}(10)$ enhancement of the signal field at the sensor position. Moreover, by tuning the bias magnetic field at the sensor, the mechanical resonance frequency can be shifted across the target band, enabling broader frequency coverage with an approximately uniform projected sensitivity. Taken together, these upgrades suggest a realistic path toward a three orders of magnitude improvement in the reach for $g_{a\gamma}$ in the sub-kHz band, highlighting the long-term potential of ferromagnet-enhanced, room-temperature platforms for ultralight dark-matter searches. With such improvements, the projected sensitivity of LeMaMa could reach $g_{a\gamma}\sim 10^{-10}$, potentially becoming competitive with, or even surpassing, existing indirect experiment laboratory bounds such as ALPS in the corresponding mass window, as indicated in Fig.\,\ref{fig:constraint}. Additional gains may be achievable through advanced low noise readout and frequency-conversion architectures~\cite{Thomson:2021zvq}, as well as quantum-enhanced measurement strategies developed in related axion and precision-sensing experiments~\cite{Martynov:2019azm}.

Beyond axion searches, the same broadband LeMaMa sensor can be adapted to probe a broad range of new physics, including electron-spin couplings, dark-photon dark matter, and exotic spin-dependent interactions \cite{Kalia:2024eml,Cong:2024qly}. More generally, its room-temperature operation, together with its high sensitivity, high stability, and broad bandwidth, makes it attractive for biomagnetic sensing, where signals from the heart, brain, and skeletal muscles typically lie in the picotesla to femtotesla range and span a broad frequency spectrum \cite{2001SuScT..14R..79P,Shah:2013eja}.
\\

\noindent \textbf{\large Methods}\\
\noindent \textbf{\large Response and noise}\\
To calibrate the detector response (i.e. the voltage on the quadrant photo detector) to an axion-converted magnetic field $B_a$, we apply a real calibration magnetic field $B$ oriented parallel to the expected axion-induced field and measure the response to this known drive.

\begin{figure*}[htb]
    \centering
\includegraphics[width=0.95\linewidth]{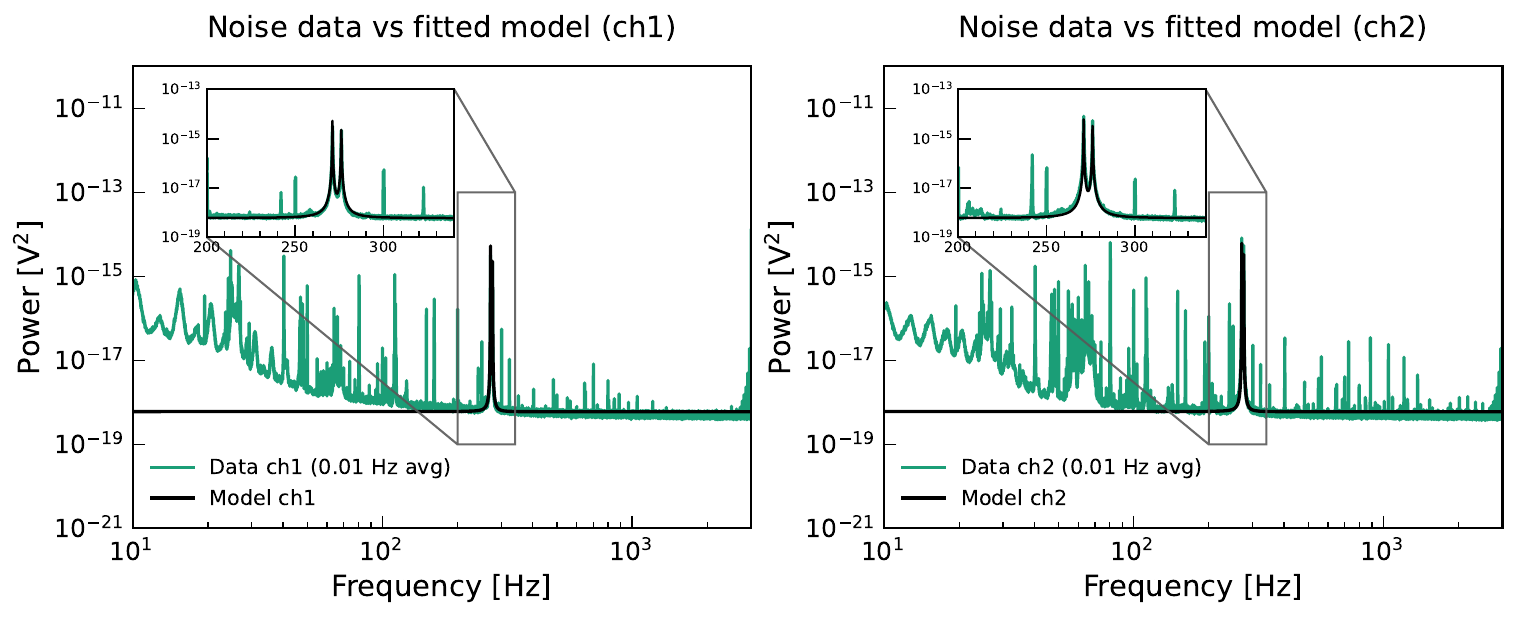}
    \caption{Measured single-channel noise spectra in the two readout channels, together with the best-fit model of Eq.~\eqref{eq:noise model}. The spectra are obtained by selecting one readout channel at a time in Eq.~\eqref{eq:noise model}: $(r_1,r_2)_{\rm ch1}=(1,0)$ for channel~1 and $(r_1,r_2)_{\rm ch2}=(0,1)$ for channel~2. The insets show the same comparison over zoomed-in frequency ranges.}
    \label{fig:noise_spectrum}
\end{figure*}

We model the detector as a two-mode linear system in a fixed modal basis.
A harmonic drive magnetic field is applied along the unit direction $\hat{\mathbf d}_1$ shown in Fig.~\ref{fig:source},
which excites two independent modes with complex Lorentzian transfer functions
$F_i(\omega)\equiv Q_i/[(\omega_i^2-\omega^2)Q_i-i\,\omega_i\,\omega] $ with $ i=1,2$.
We also define the orthogonal input direction $\hat{\mathbf d}_2$,
which will be used to describe coupled noise entering transverse to the signal direction.
The quadrant detector provides two readout channels corresponding to the (orthonormal) detection directions $\hat{\mathbf m}_1$ and $\hat{\mathbf m}_2$.

We describe the detector response by a complex transfer matrix $H_{ij}(\omega)$ in Eq.~\eqref{eq:fit-model}, where $j\in\{1,2\}$ labels the magnetic-field input direction $\hat{\mathbf d}_j$ and $i\in\{1,2\}$ labels the optical readout channel defined by $\hat{\mathbf m}_i$. 
Thus, $H_{ij}$ maps an input field along $\hat{\mathbf d}_j$ to the voltage response in channel $i$, with units of $\mathrm{V/T}$. The calibration data constrain the magnitude responses
$|H_{ij}(\omega)|$.
In the calibration run, we inject the drive field along $\hat{\mathbf d}_1$ and fit the measured responses in both readout channels, $|H_{11}(\omega)|$ and $|H_{21}(\omega)|$, to the model above.
The fitted response curves are shown in Figs.~\ref{fig:response}. The best-fit values of the eight parameters are summarized in Table~\ref{tab:response_params}.

For the axion search, we combine the readouts from two directions $\hat{\mathbf m}_1$ and $\hat{\mathbf m}_2$ in frequency domain. The weights for axion DM are chosen to be 
\begin{align}
 (r_1,r_2)_{\rm dm}\equiv\frac{1}{\sqrt{H_{11}^2+H_{21}^2}}(H_{11}^*,H_{21}^*).
\end{align}
Therefore, the resulting effective response for signal from direction $\hat{\mathbf d}_1$ is
\begin{equation}\label{eq:eta}
\eta(\omega)\equiv \sqrt{|H_{11}(\omega)|^2+|H_{21}(\omega)|^2}.
\end{equation}

We work with the discrete Fourier transform (DFT) of the measured time series data over an observation time $T=N\Delta t$~\cite{Lee:2022vvb},
\begin{equation}
\tilde{\beta}_k=\sum_{n=0}^{N-1}\beta_n\,e^{-i\omega_k n\Delta t},
\qquad
\omega_k=\frac{2\pi k}{N\Delta t},
\end{equation}
where $k$ is the index of the frequency, and we define the real quadratures
\begin{equation}\label{eq:FFT_quadratures_SM}
A_k=\frac{2}{N}\,\mathrm{Re}\,\tilde{\beta}_k,
\qquad
B_k=-\frac{2}{N}\,\mathrm{Im}\,\tilde{\beta}_k.
\end{equation}

We model the noise as the sum of an uncoupled white component and a coupled white component.
The uncoupled noise enters directly in each readout channel with variance $\sigma_{\rm unc}^2$ and is assumed to be uncorrelated between channels.
The coupled noise is taken to be Gaussian, stationary, and white in the two orthogonal input directions $\hat{\mathbf d}_1$ and $\hat{\mathbf d}_2$, with a common strength $\sigma_c^2$.
After combination, the coupled noise is shaped by the complex transfer functions.
For the combined readout of two channels $\hat{\mathbf m}_1$ and $\hat{\mathbf m}_2$ with unassigned weights $(r_1,r_2)$, the covariance of the FFT quadratures satisfies
\begin{align}
\mathrm{Cov}&(A_k,A_r)=\mathrm{Cov}(B_k,B_r) \label{eq:noise model} \\
=&\frac{2}{N}\big[
\sigma_{\rm unc}^2+\sigma_c^2\,\bigl|r_1H_{11}(\omega_k)+r_2H_{21}(\omega_k)\bigr|^2
\nonumber \\
&+\sigma_c^2\,\bigl|r_1H_{12}(\omega_k)+r_2H_{22}(\omega_k)\bigr|^2
\big]\delta_{kr}, \nonumber \\
=&\frac{2}{N}\left[
\sigma_{\rm unc}^2
+\sigma_c^2 \eta^2_N(\omega)
	\right]\delta_{kr}, \nonumber
\end{align}
where $N$ is the number of time samples and $(A_k,B_k)$ are the cosine/sine (real/imaginary) FFT quadratures at angular frequency $\omega_k$.
Here we define the effective response curve for coupled-noise as
\begin{align}
\eta_N(\omega)& \equiv
\bigl|r_1H_{11}(\omega)+r_2H_{21}(\omega)\bigr|^2 \nonumber \\
& +\bigl|r_1H_{12}(\omega)+r_2H_{22}(\omega)\bigr|^2.
\label{eq:eta_bar}
\end{align}

\begin{table}[htb]
\centering
\caption{Best-fit noise-model parameters for Eq.~\eqref{eq:noise model}, obtained from a joint fit to the two single-channel noise spectra shown in Fig.~\ref{fig:noise_spectrum}. Here $\sigma_{\rm unc}$ is the per-channel uncoupled white-noise level (uncorrelated between channels), expressed in the native readout (voltage) units. The parameter $\sigma_c$ is the common coupled white-noise strength defined in the orthogonal input basis $(\hat{\mathbf d}_1,\hat{\mathbf d}_2)$; its unit is that of the effective input (e.g., T if $H_{ij}$ is in V/T).}
\renewcommand{\arraystretch}{1.15}
\begin{tabular}{l c c}
\hline\hline
Parameter & Best-fit value & Unit \\
\hline
$\sigma_{\rm unc}$ & $9.003\times 10^{-6}$ & V \\
$\sigma_{c}$ & $1.705\times 10^{-11}$ & T \\
\hline\hline
\end{tabular} \label{tab:noise_params}
\end{table}

In Fig.~\ref{fig:noise_spectrum}, we show the measured noise spectra for the two individual readout channels, corresponding to the special cases of Eq.~\eqref{eq:noise model} with fixed weights $(r_1,r_2)_{\rm ch1}=(1,0)$ and $(r_1,r_2)_{\rm ch2}=(0,1)$. We fit both spectra with the same noise model, using common parameters $\sigma_{\rm unc}$ and $\sigma_c$, and report their best-fit values in Table~\ref{tab:noise_params}.

The model reproduces the main resonant feature and the overall broadband noise level reasonably well, especially near the resonance region. 
Residual discrepancies are most evident at low frequencies, where broad excess noise and isolated narrow lines are observed. The broad excess is consistent with vibration-induced motion, whereas the narrow features near 50 Hz harmonics and an approximately 80 Hz-spaced line family likely originate from power-line pickup and the National Instruments acquisition chassis, leading to locally weaker limits.

With the fitted response and measured noise with weight $r_{\rm dm}$, the magnetic-field resolution $R(\omega)$ and the magnetic-field sensitivity $S(\omega)$ are related as
\begin{equation}
R(\omega)\equiv\frac{1}{\sqrt{T}}S(\omega)\equiv\frac{1}{\eta(\omega)\sqrt{T}}\sqrt{\mathrm{Cov}(A_\omega,A_\omega)_{r_{\mathrm{dm}}}},
\label{eq:S_omega}
\end{equation}
The resulting magnetic-field resolution is shown in Fig.~\ref{fig:resolution}.
\\
The experiment-specific transduction from the axion field to the effective magnetic observable is fully encoded in $\mathcal{K}$, as shown in Eq.\,\eqref{eq:Ba_omega}.  With the stochastic axion field described in Eq.~\eqref{eq:axionfield}, the variance per bin can be written as 
\begin{align}
\label{eq:Sigma_a_SM}
\Sigma_a(\omega_k;g_{a\gamma},m_a)
& =g_{a\gamma}^2
\mathcal{K}^2\,
\frac{8\pi^2\rho_{\rm DM}}{m_a\,T}\,
\frac{1}{(2\pi\sigma_v^2)^{3/2}} \\
& \times \exp\!\left[-\frac{v_k^2+v_E^2}{2\sigma_v^2}\right]\,
v_k\,\frac{\sinh b_k}{b_k}, \nonumber 
\end{align}
where $\sqrt{2}\,\sigma_v=v_0=220~\mathrm{km/s}$ is the velocity dispersion, and $\mathbf{v}_E$ is the Earth's velocity in the Galactic frame~\cite{Bovy:2009dr}, $T$ represents measurement time, and
\begin{equation}
v_k=\sqrt{2\left(\frac{\omega_k}{m_a}-1\right)},
\qquad
b_k\equiv \frac{v_k v_E}{\sigma_v^2}.
\end{equation}
Here $\Sigma_a$ is the variance of the effective magnetic-field quadratures $(A_k,B_k)$ in each Fourier bin and therefore carries units of $\mathrm{T}^2$.

For the axion signal alone, $(A_k,B_k)$ are jointly Gaussian with vanishing mean.
For sufficiently long $T$ (i.e., when the measurement time $T$ is much longer than the axion coherence time $\tau_a$ in our case), the finite-time sinc kernel appearing in the DFT of a narrowband signal can be approximated by a Dirac delta,
\begin{equation}
\frac{\sin\!\left(\tfrac{T}{2}(\omega-\omega_k)\right)}{\omega-\omega_k}\ \longrightarrow\ \pi\,\delta(\omega-\omega_k),
\end{equation}
which renders different Fourier bins approximately uncorrelated.
Under this delta-function approximation, the signal covariance takes the diagonal form
\begin{align}
\mathrm{Cov}(A_k,A_r) & =\mathrm{Cov}(B_k,B_r)=\delta_{kr}\,\Sigma_a(\omega_k), \label{eq:axion_cov_diag_SM} \\
\mathrm{Cov}(A_k,B_r) & =0. 
\end{align} 

The axion induced effective magnetic field covariance matrix contracted using Eq.~\eqref{eq:Sigma_a_SM} can be converted to voltage by multiplying $\eta(\omega)$ obtained in Eq.~\eqref{eq:eta}.
Meanwhile, the measured frequency-domain noise is modeled using Eq.~\eqref{eq:noise model} with  $(r_1,r_2)_{\rm dm}$. 
The likelihood analysis for each axion mass is then performed by comparing these two quantities, from which the corresponding coupling constant can be extracted at each frequency, with noise parameters $\sigma_{\rm unc}$ and $\sigma_{\rm c}$ as nuisance parameters.
The full likelihood and the profile-likelihood-ratio framework are summarized in Supplementary Note 1.
\\

\begin{figure}[htb]
    \centering
        \includegraphics[width=0.95\linewidth]{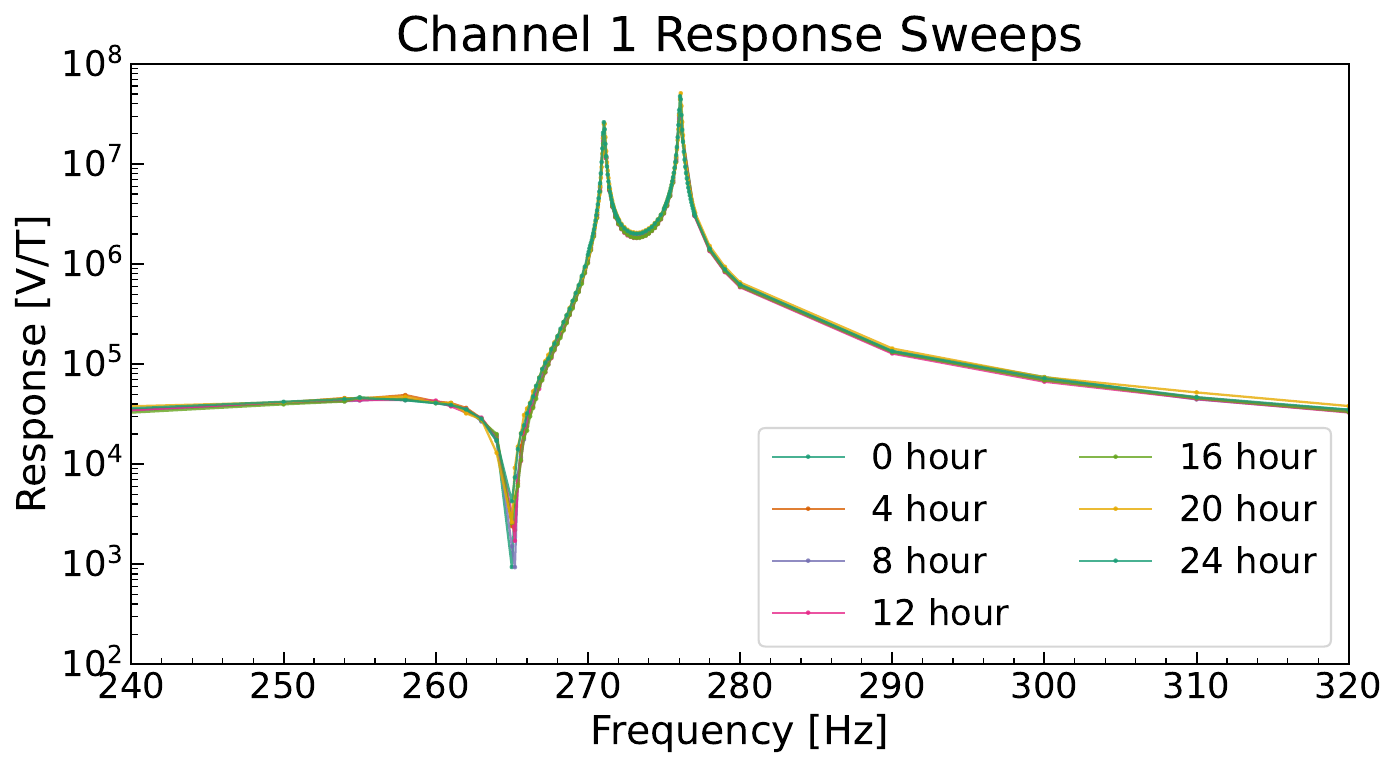} \\
        \includegraphics[width=0.95\linewidth]{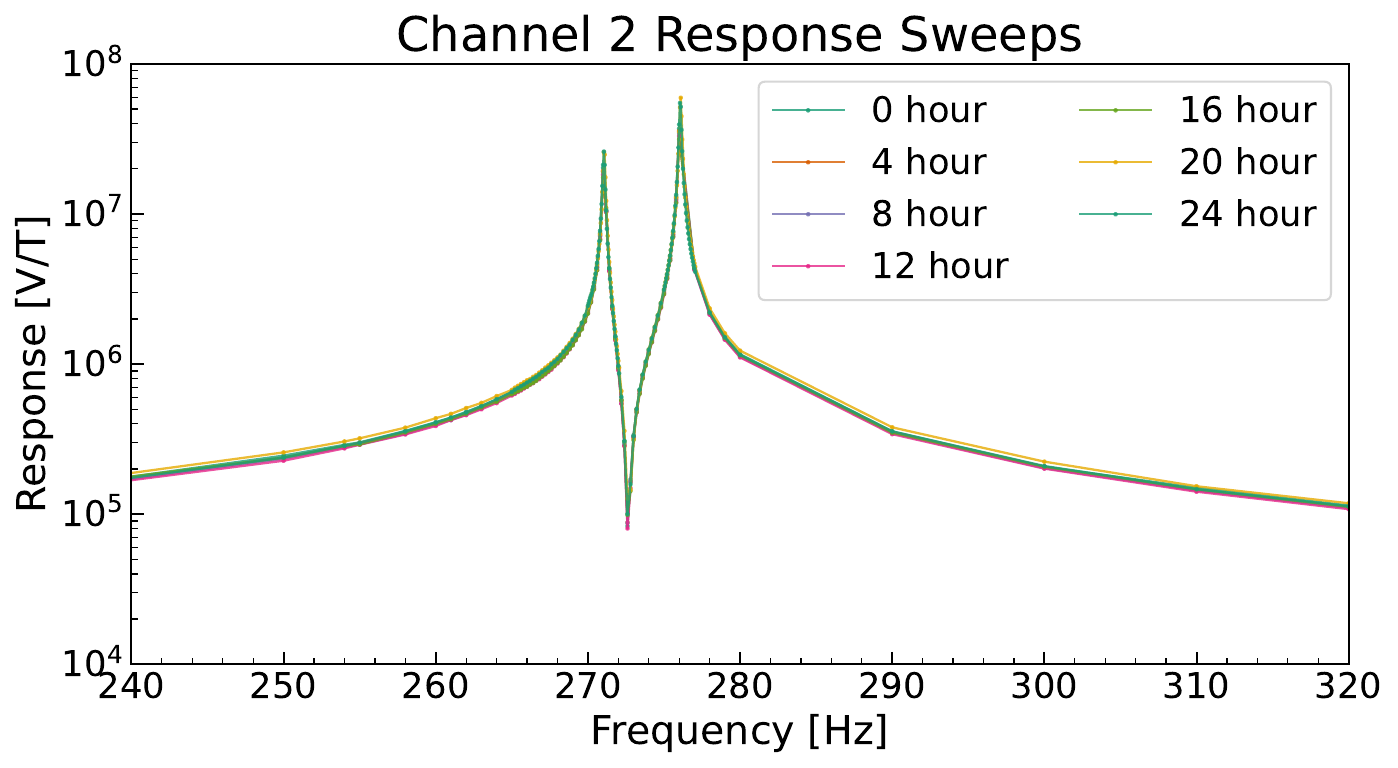}
    \caption{Repeated response sweeps taken every $4$~h over a $24$~h period (seven scans total). Left: channel~1; right: channel~2. The overlaid curves show that the resonance structure, including the narrow antiresonance feature, is stable across scans.}
    \label{fig:response_stability}
\end{figure}

\noindent \textbf{\large Response stability}\\
To assess the long-term stability of the calibrated transfer functions used in the likelihood analysis, we repeated the response sweep every $4$~h over a $24$~h period (seven scans in total).
As shown in Fig.~\ref{fig:response_stability}, the resonance structure in both readout channels is highly reproducible: the peak locations and the narrow antiresonance feature remain aligned across all scans.

\begin{figure}[htb]
    \centering
    \includegraphics[width=0.95\linewidth]{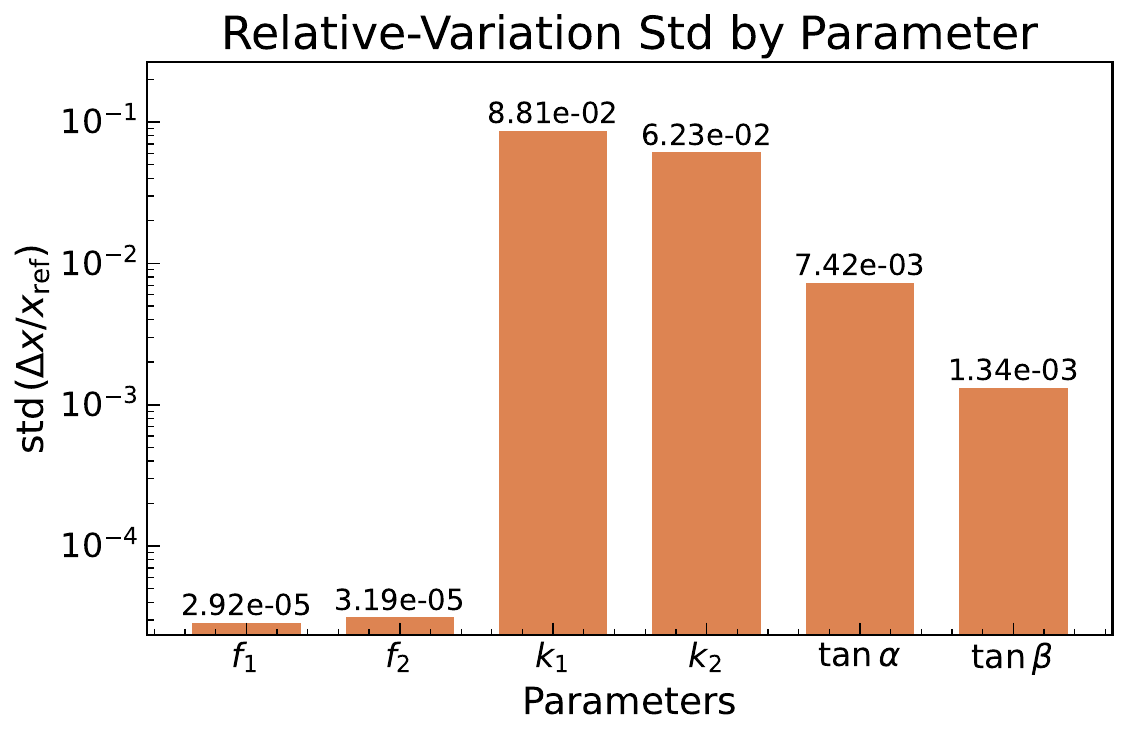}
    \caption{Relative standard deviation of the fitted response-model parameters across the seven sweeps. The resonant frequencies are stable at the $\mathcal{O}(10^{-5})$ level, corresponding to sub-$0.02$~Hz absolute drift near $270$--$276$~Hz, while variations in the mode weights and overall gains primarily rescale the response amplitude.}
    \label{fig:response_stability_params}
\end{figure}

Fitting each sweep with the same two-mode model yields the parameter variations summarized in Fig.~\ref{fig:response_stability_params}.
In particular, the fractional standard deviations of the resonant frequencies are at the level of $\mathcal{O}(10^{-5})$, implying an absolute drift well below $0.02$~Hz near the resonance band.
This is much smaller than the $\sim 3$~Hz-wide region where the magnetic resolution satisfies $R(\omega)\lesssim 1~\mathrm{fT}/\sqrt{\mathrm{Hz}}$ (see Fig.~\ref{fig:resolution}), so the observed response drift is negligible compared with the effective bandwidth that dominates our resolution.
These measurements support that the calibrated response is stable over the integration time used for the axion search.

\begin{acknowledgments}
The authors would like to thank Xiaoping Wang and Jinhui Guo for valuable discussions. This work is supported by the National Science Foundation of China under Grant No. 12235001, No. 12475103 and State Key Laboratory of Nuclear Physics and Technology under Grant No. NPT2025ZX07 and No. NPT2025ZX11, and “The Fundamental Research Funds for the Central Universities, Peking University”. This work was also supported by the Cluster of Excellence ``Precision Physics, Fundamental Interactions, and Structure of Matter'' (PRISMA++ EXC 2118/2) funded by the German Research Foundation (DFG) within the German Excellence Strategy (Project ID 390831469) and by the COST Action within the project COSMIC WISPers (Grant No.\ CA21106).
\end{acknowledgments}

\clearpage
\onecolumngrid
\renewcommand{\figurename}{FIG S.}
\renewcommand{\tablename}{TABLE S.}
\setcounter{figure}{0}
\setcounter{table}{0}
\renewcommand{\theHfigure}{S.\arabic{figure}}
\renewcommand{\theHtable}{S.\arabic{table}}

\section{Supplementary Note 1: Log-Likelihood-Ratio  analysis}\label{sec:appendix_LLR}

\subsection{Profile-likelihood framework and test statistics}

We use a log-likelihood-ratio (LLR) analysis to test the background-only hypothesis and to set upper limits on the axion--photon coupling $g_{a\gamma}$~\cite{Cowan:2010js}.
For each scanned axion mass point, we work with the real FFT quadratures in the combined readout (in volts),
\begin{equation}
\bm{d}\equiv\{A_k,B_k\}_{k=0}^{N-1},
\end{equation}
which are modeled as a zero-mean multivariate normal distribution with diagonal covariance.
The axion contribution is taken to be Gaussian with diagonal covariance
\begin{equation}
\hat{\bm{\Sigma}}_a(\omega;g_{a\gamma},m_a)=\Sigma_a(\omega;g_{a\gamma},m_a)\,\eta^2(\omega),
\end{equation}
where $\eta(\omega)$ is the calibrated frequency response (with units of $\mathrm{V}/\mathrm{T}$) and $\Sigma_a$ is the intrinsic axion spectral variance in magnetic-field units (with units of $\mathrm{T}^2$).
Accordingly, $\hat{\bm{\Sigma}}_a$ carries units of $\mathrm{V}^2$.

The background is modeled as the sum of two independent white-noise components: an uncoupled contribution and a coupled contribution shaped by the calibrated transfer functions after combination (see Response and noise section in Methods).
The total covariance for each frequency bin can be written compactly as
\begin{equation}
\bm{\Sigma}(\omega)
=
\hat{\bm{\Sigma}}_a(\omega;g_{a\gamma},m_a)
+\bm{\Sigma}_b(\omega),
\qquad
\bm{\Sigma}_b(\omega)\equiv \frac{2}{N}\Big[\sigma_{\rm unc}^2+\sigma_c^2\,\bar{\eta}(\omega)\Big]\bm{I},
\label{eq:total_covariance_compact}
\end{equation}
where $\bm{I}$ is the $2\times 2$ identity matrix acting on $(A_k,B_k)$ and $\bar{\eta}(\omega)$ is defined in Methods.
The likelihood is then
\begin{equation}
L(\bm{d}\,|\,g_{a\gamma}^2,\sigma_{\rm unc}^2,\sigma_c^2)
=
\frac{1}{\sqrt{(2\pi)^{2N}\det\bm{\Sigma}}}
\exp\!\left(-\frac{1}{2}\bm{d}^{\mathsf T}\bm{\Sigma}^{-1}\bm{d}\right).
\label{eq:gaussian_likelihood_compact}
\end{equation}

To set an upper limit we parameterize the non-negative signal strength as $\mu\equiv g_{a\gamma}^2$ and profile over the nuisance parameters
$\theta\equiv(\sigma_{\rm unc}^2,\sigma_c^2)$.
We use the one-sided profile-likelihood-ratio statistic
\begin{equation}
q(\mu)=
\begin{cases}
2\left[\ell(\hat{\mu},\hat{\theta})-\ell\!\left(\mu,\hat{\hat{\theta}}(\mu)\right)\right], & \hat{\mu}\le \mu,\\[4pt]
0, & \hat{\mu}>\mu,
\end{cases}
\label{eq:qmu_def_compact}
\end{equation}
where hats denote unconditional maximum-likelihood estimates and double hats denote conditional maximization at fixed $\mu$.
For discovery we use the special case $q_0\equiv q(\mu=0)$.
In the asymptotic regime~\cite{Cowan:2010js}, $q(\mu)$ follows the usual half-$\chi^2$ mixture distribution under $\mu=\mu^\ast$,
\begin{equation}
q(\mu)\ \dot{\sim}\ \frac{1}{2}\delta(q)+\frac{1}{2}\chi^2_1(q),
\label{eq:half_chi2_compact}
\end{equation}
so that the local significance is $Z\simeq\sqrt{q_0}$.

If a discovery candidate were to be established, we also construct a two-sided confidence interval for the signal-strength parameter $\mu$ using the likelihood-ratio statistic
\begin{equation}
t(\mu)\equiv 2\left[\ell(\hat{\mu},\hat{\theta})-\ell\!\left(\mu,\hat{\hat{\theta}}(\mu)\right)\right],
\label{eq:tmu_def}
\end{equation}
i.e.\ without setting the statistic to zero when $\hat{\mu}>\mu$.
In the asymptotic limit, $t(\mu)$ follows a $\chi^2$ distribution with one degree of freedom under $\mu=\mu^\ast$~\cite{Cowan:2010js}, so the two-sided $95\%$ confidence interval for $\mu$ is obtained by solving
\begin{equation}
t(\mu)=t_{\mathrm{crit}},
\qquad
t_{\mathrm{crit}}=\chi^2_{1;0.95}\simeq 3.84,
\label{eq:tmu_crit}
\end{equation}
and mapping the resulting interval $[\mu_{\rm low},\mu_{\rm high}]$ to
$g_{a\gamma}\in\big[\sqrt{\mu_{\rm low}},\sqrt{\mu_{\rm high}}\big]$.
This two-sided construction is used when interpreting injection-and-recovery tests, where the recovered best-fit couplings should lie within the nominal discovery interval at the injected frequency.

\begin{figure}[tpb]
    \centering
    \includegraphics[width=1\linewidth]{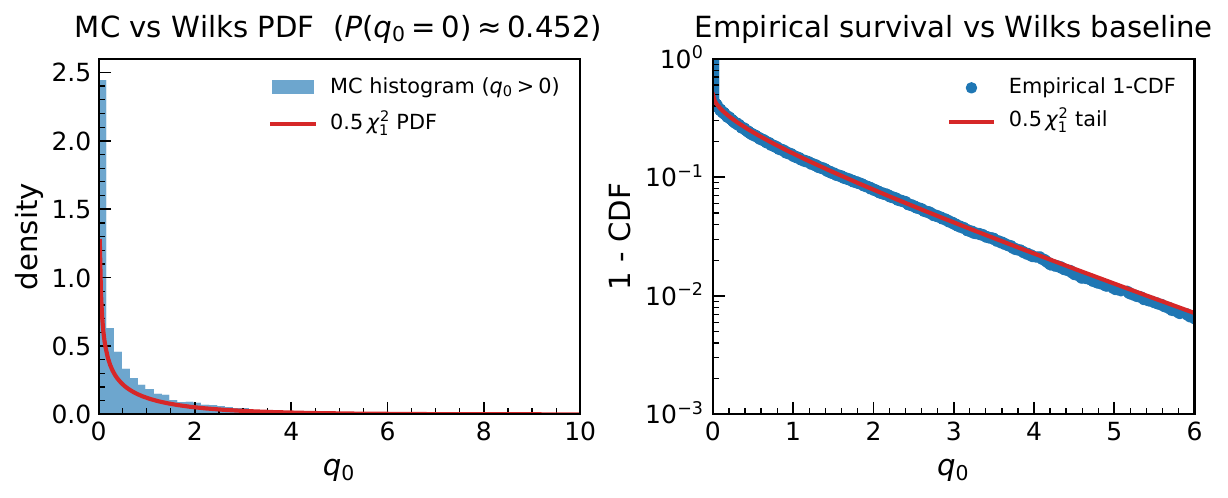}
    \caption{Monte Carlo validation of the asymptotic null distribution of the discovery statistic $q_0$ at a representative frequency ($370~\mathrm{Hz}$). Left: histogram of Monte Carlo samples with $q_0>0$, compared to the Wilks prediction for the continuous component, $\tfrac{1}{2}\chi^2_1$ (red curve). The fraction of trials with $q_0=0$  is $P(q_0=0)\simeq 0.452$, consistent with the expected point mass near $\tfrac{1}{2}$ from the one-sided definition. Right: empirical survival function $1-\mathrm{CDF}$ of $q_0$ compared to the predicted $\tfrac{1}{2}\chi^2_1$ tail, showing good agreement over the range relevant for local $p$-values.}
    \label{fig:Wilks}
\end{figure}

We validate the asymptotic null approximation for $q_0$ using Monte Carlo simulations generated with the measured response and noise parameters.
Figure~\ref{fig:Wilks} shows the results at $370~\mathrm{Hz}$: the $q_0>0$ density follows the predicted $\tfrac{1}{2}\chi^2_1$ form, and the survival function agrees with the corresponding tail, supporting the use of the asymptotic mapping between $q_0$ and local significance.

\subsection{Look-elsewhere effect and global discovery threshold}\label{subsec:appendix_LEE}

\begin{figure}[tpb]
    \centering
    \begin{minipage}[t]{0.49\linewidth}
        \vspace{0pt}
        \centering
        \includegraphics[width=\linewidth]{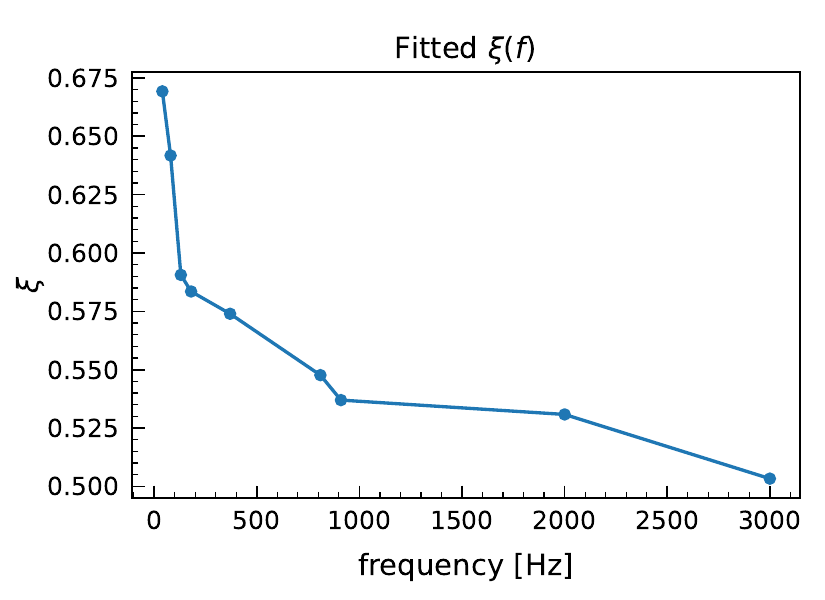}
        \caption{Fitted correlation parameter $\xi(f)$ that maps the axion linewidth to an effective number of independent trials. The fit is performed in several sub-intervals across the scan band, and the resulting $\xi$ values are interpolated as a function of the scan frequency.}
        \label{fig:alpha}
    \end{minipage}\hfill
    \begin{minipage}[t]{0.49\linewidth}
        \vspace{0pt}
        \centering
        \includegraphics[width=\linewidth]{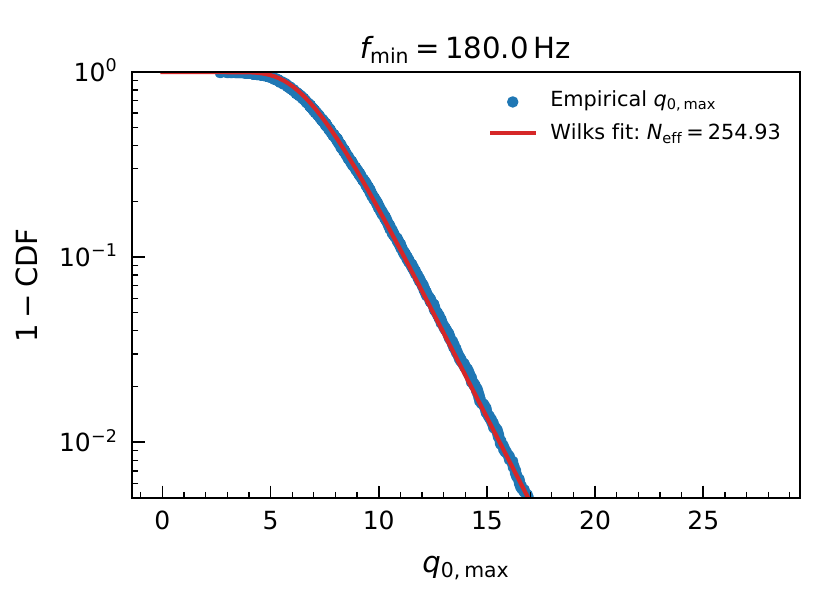}
        \caption{Example look-elsewhere calibration in a representative scan sub-interval starting at $180~\mathrm{Hz}$. The Monte Carlo distribution of the maximum discovery statistic $q_{\mathrm{th}}=\max(q_0)$ in this sub-interval is fitted to extract the effective number of independent trials. For this interval we obtain $N=254.93$.}
        \label{fig:LookElseWhere}
    \end{minipage}
\end{figure}

We adopt the conventional discovery criterion of a global significance of $5\sigma$ for a blind scan over many correlated mass hypotheses.
Concretely, we define
\begin{equation}
p_{5\sigma}\equiv 1-\Phi(5)\simeq 2.87\times 10^{-7},
\end{equation}
and require that the global tail probability for obtaining a fluctuation at least as significant anywhere in the scanned band satisfies
\begin{equation}
p_{\rm global}(q_{\rm th}) = p_{5\sigma}.
\label{eq:global5sigma_def}
\end{equation}
Using the one-sided asymptotic mapping $p_{\rm local}(q)=1-\Phi(\sqrt{q})$ and the effective number of independent trials $N_{\mathcal F}$, we approximate
\begin{equation}
p_{\rm global}(q)=1-\bigl(1-p_{\rm local}(q)\bigr)^{N_{\mathcal F}}
=1-\bigl(1-[1-\Phi(\sqrt{q})]\bigr)^{N_{\mathcal F}}.
\label{eq:pglobal_from_plocal}
\end{equation}
A discovery candidate is flagged if $\max(q_0)>q_{\rm th}$; otherwise the data are treated as compatible with the background-only hypothesis and we proceed to set upper limits.

We scan over axion frequencies on a grid with spacing $\Delta f_a/2$~\cite{Lee:2022vvb}, where $\Delta f_a$ denotes the characteristic axion linewidth in the Standard Halo Model (SHM), set by virial broadening,
\begin{equation}
\Delta f_a \equiv f_a\,v_0^2,
\qquad
f_a=\frac{m_a}{2\pi},
\end{equation}
with $v_0\simeq 220~\mathrm{km/s}$ the characteristic halo speed.
For each tested axion frequency $f_a$, the profile likelihood is evaluated only over the subset of FFT bins satisfying
\begin{equation}
f_k \in
\left[
f_a-0.2\,\delta_{\rm win}f_a,\;
f_a+\delta_{\rm win}f_a
\right],
\qquad
\delta_{\rm win}=6\times10^{-6}.
\label{eq:llr_window_def}
\end{equation}
This choice is intentionally asymmetric. In our signal model, the Standard Halo Model axion line is supported predominantly on the $f\gtrsim f_a$ side, as also reflected in the expression for $\Sigma_a$ in Methods through
\begin{equation}
v_k=\sqrt{2\left(\frac{\omega_k}{m_a}-1\right)}.
\end{equation}
We therefore retain a larger high-frequency sideband to capture the virial-broadened signal tail, while keeping a small low-frequency margin.
The choice of scan spacing $\Delta f_a/2$ ensures that neighboring mass hypotheses overlap within one linewidth, while the above analysis window provides near-optimal coverage of the narrow axion spectral feature.
Because adjacent scan points probe highly correlated data, the relevant discovery threshold must be set by the global significance rather than the local one (the look-elsewhere effect).
Following Ref.~\cite{Lee:2022vvb}, we parameterize the correlation length by introducing a factor $\xi$ such that hypotheses separated by a fractional step of order $\xi v_0^2$ are approximately independent.
Equivalently, an approximately logarithmic grid of independent masses can be written as
\begin{equation}
m_a^{(i)} = m_a^{(0)}\bigl(1+\xi v_0^2\bigr)^i,
\end{equation}
so that, for a frequency interval $[f_{\min},f_{\max}]$ with $\xi v_0^2\ll 1$, the effective number of independent trials is
\begin{equation}
N \simeq \frac{1}{\xi v_0^2}\,\ln\!\left(\frac{f_{\max}}{f_{\min}}\right).
\label{eq:N_alpha_relation}
\end{equation}

In practice, we determine $\xi$ empirically using null Monte Carlo simulations.
We divide the full scan band into several short sub-intervals $[f_{\min,i},f_{\max,i}]$, generate an ensemble of background-only datasets using the measured response and noise model, and run the full scan within each sub-interval using the same $\Delta f_a/2$ grid as in the data analysis.
Concretely, we perform the look-elsewhere calibration in nine representative sub-intervals whose lower edges are
\begin{equation}
f_{\min,i}\in\{40,\,80,\,130,\,180,\,370,\,810,\,910,\,2000,\,3000\}~\mathrm{Hz},
\end{equation}
and for each sub-interval we choose a fixed fractional width
\begin{equation}
\frac{f_{\max,i}}{f_{\min,i}}=1.00008,
\end{equation}
so that the interval is narrow in $\ln f$ while still containing many scanned frequency points.

\begin{table}[tpb]
\centering
\caption{Look-elsewhere calibration sub-intervals and extracted effective numbers of independent trials. For each sub-interval $[f_{\min,i},f_{\max,i}]$ with $f_{\max,i}/f_{\min,i}=1.00008$, we fit the Monte Carlo distribution of $q_{\mathrm{th}}=\max(q_0)$ to extract $N_i$ and infer the corresponding correlation parameter $\xi(f_{\min,i})$ via Eq.~\eqref{eq:N_alpha_relation}.}
\label{tab:LEE_points}
\renewcommand{\arraystretch}{1.15}
\begin{tabular}{c c c c c}
\hline\hline
$f_{\min,i}$ [Hz] & $f_{\max,i}$ [Hz] & $\xi(f_{\min,i})$ & $N_i$ & MC samples \\
\hline
$40.002$   & $40.005$   & $0.6692$ & $222.28$ & $10000$ \\
$80.003$   & $80.010$   & $0.6417$ & $231.80$ & $10000$ \\
$130.005$  & $130.016$  & $0.5906$ & $251.87$ & $10001$ \\
$180.007$  & $180.022$  & $0.5835$ & $254.93$ & $10001$ \\
$370.015$  & $370.044$  & $0.5740$ & $259.17$ & $10000$ \\
$810.032$  & $810.097$  & $0.5477$ & $271.61$ & $9999$ \\
$910.036$  & $910.109$  & $0.5370$ & $277.02$ & $9999$ \\
$2000.080$ & $2000.240$ & $0.5308$ & $280.22$ & $10000$ \\
$3000.120$ & $3000.360$ & $0.5033$ & $295.53$ & $10000$ \\
\hline\hline
\end{tabular}
\end{table}

For each Monte Carlo realization we record the maximum discovery statistic,
\begin{equation}
q_{\mathrm{th}} \equiv \max(q_0),
\end{equation}
over that sub-interval.
If $P_{\mathrm{local}}(q)$ denotes the local null tail probability for a single test (given by the half-$\chi^2$ form in Eq.~\eqref{eq:half_chi2_compact}), then the global tail probability for the maximum in an interval with $N$ effective independent trials is
\begin{equation}
P_{\mathrm{global}}(q)=1-\bigl(1-P_{\mathrm{local}}(q)\bigr)^N.
\label{eq:Pglobal_fit}
\end{equation}
We fit Eq.~\eqref{eq:Pglobal_fit} to the Monte Carlo distribution of $q_{\mathrm{th}}$ to extract $N$ for each sub-interval, and then infer $\xi$ via Eq.~\eqref{eq:N_alpha_relation}.
Figure~\ref{fig:LookElseWhere} shows an example fit in the sub-interval starting at $180~\mathrm{Hz}$, for which we obtain $N=254.93$.
Repeating this procedure across the scan band yields a set of $\xi$ values, which we interpolate to obtain $\xi(f)$ (Fig.~\ref{fig:alpha}).

Finally, we combine the frequency-dependent correlation calibration to estimate the total effective number of independent trials across the scan band $\mathcal{F}$ used in the analysis.
In this work, the global-trials integral is evaluated over $\mathcal{F}\equiv[40,\,3000]~\mathrm{Hz}$ (after fitting/interpolating $\xi(f)$ using the calibration points above), via
\begin{equation}
N_{\mathcal{F}}
= \int_{\ln\!\bigl(\min(\mathcal{F})\bigr)}^{\ln\!\bigl(\max(\mathcal{F})\bigr)}
\frac{1}{\xi(\nu)\,v_0^2}\,d\nu,
\qquad \nu\equiv \ln f,
\end{equation}
and use the resulting $N_{\mathcal{F}}$ to set a global discovery threshold for the maximum $q_0$ across the full scan.
In our analysis we obtain $N_{\mathcal{F}}=1.4\times 10^7$ and adopt a global discovery threshold $q_{\rm th}=55.76$.

\subsection{Upper-limit setting and injection recovery}
For limit setting, we first require that the data are compatible with the background-only hypothesis,
\begin{equation}
q_0<q_{\rm th}.
\end{equation}
In this case we set a one-sided $95\%$ confidence-level upper limit on the non-negative signal-strength parameter $\mu\equiv g_{a\gamma}^2$ using the profile-likelihood-ratio statistic $q(\mu)$ in Eq.~\eqref{eq:qmu_def_compact}.
The upper limit $\mu_{95}$ is defined by
\begin{equation}
q(\mu_{95})=\big[\Phi^{-1}(0.95)\big]^2\simeq 2.71,
\qquad
g_{a\gamma}^{95}=\sqrt{\mu_{95}},
\label{eq:mu95_def}
\end{equation}
where $\Phi$ is the standard normal cumulative distribution function.

\begin{figure}[!tpb]
    \centering
    \includegraphics[width=0.7\linewidth]{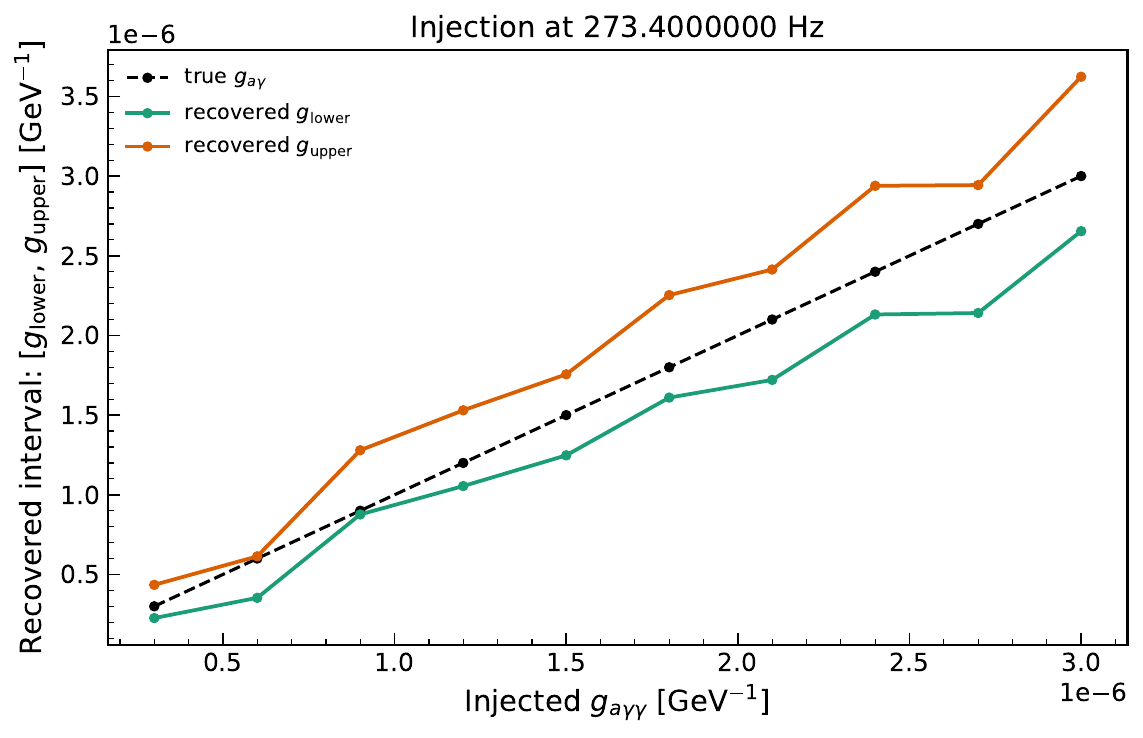}
    \caption{Injection-and-recovery validation of the LLR analysis at $f_a=270.4~\mathrm{Hz}$.}
    \label{fig:Inject}
\end{figure}

We validate the end-to-end pipeline by injecting simulated axionlike signals into the measured time series and re-running the full analysis.
As a representative example, we inject signals at $f_a=270.4~\mathrm{Hz}$ with $g_{a\gamma}\in[3\times10^{-7},\,3\times10^{-6}]~\mathrm{GeV}^{-1}$ and confirm that the recovered best-fit couplings are consistent with the injected values (Fig.~\ref{fig:Inject}).

\section{Supplementary Note 2: Recurrent candidate outliers in validation datasets}
\label{sec:appendix_recurrent_outliers}
Among the 106 narrowband candidate outliers identified in Dataset~3, 53 were
also found with comparable significance in the independent validation datasets,
Datasets~4 and 5. These recurrent candidates cluster into five narrow frequency
intervals. To examine their physical origin, we plot representative spectra in
these five intervals, choosing in each interval the candidate with the largest
test statistic $q_0$. The comparison shows a clear shape mismatch between the
axion-model spectrum and the observed features: the measured lines are
significantly narrower than the expected axion spectral profile under the
Standard Halo Model, and are instead consistent with narrow instrumental-noise
structures. Therefore, these recurrent high-significance candidates are
interpreted as device-intrinsic noise rather than axion signals.

\begin{figure*}[t]
\centering
\subfloat[]{
\includegraphics[width=0.32\textwidth]{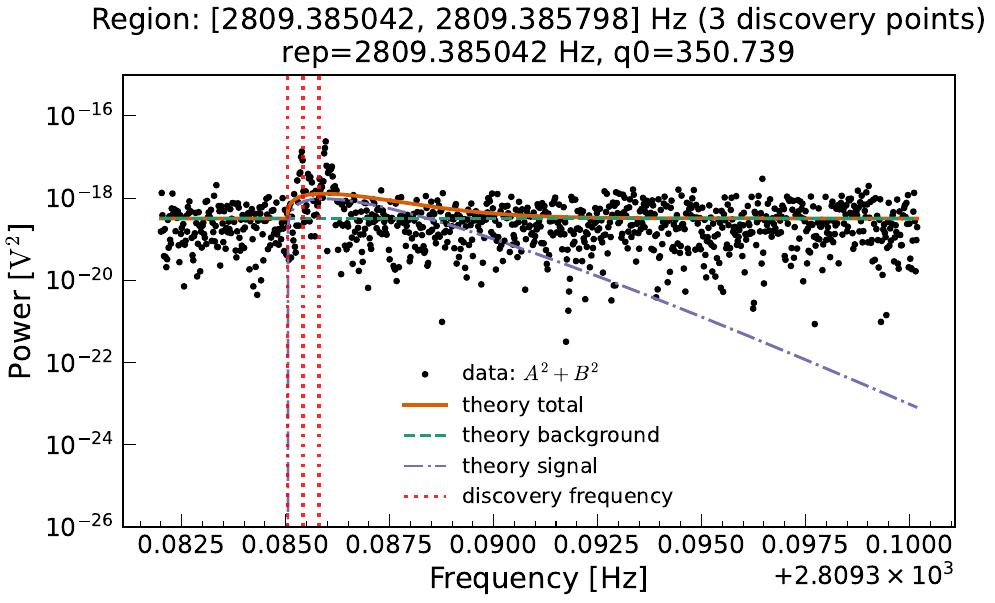}}
\hfill
\subfloat[]{
\includegraphics[width=0.32\textwidth]{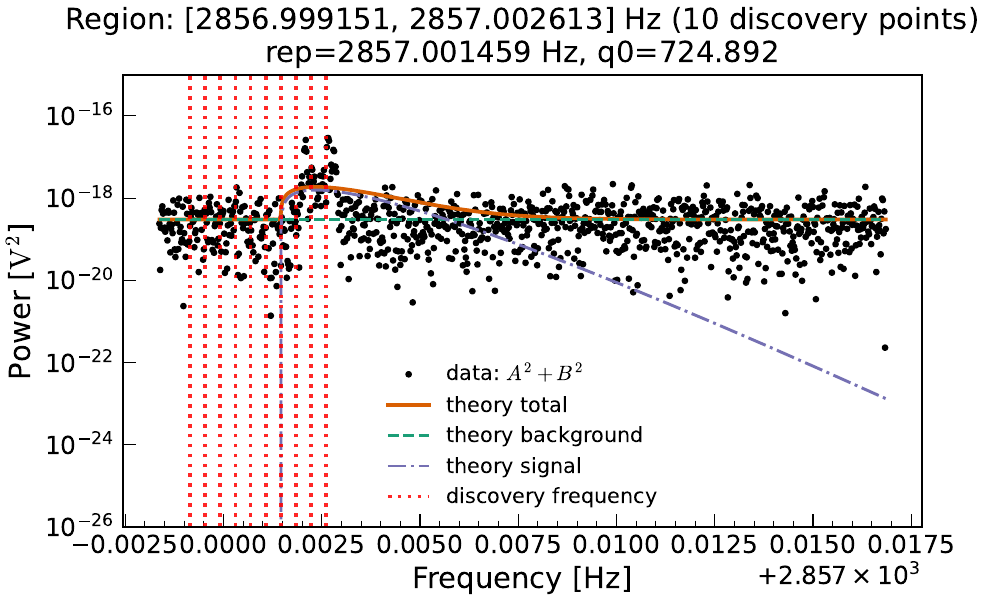}}
\hfill
\subfloat[]{
\includegraphics[width=0.32\textwidth]{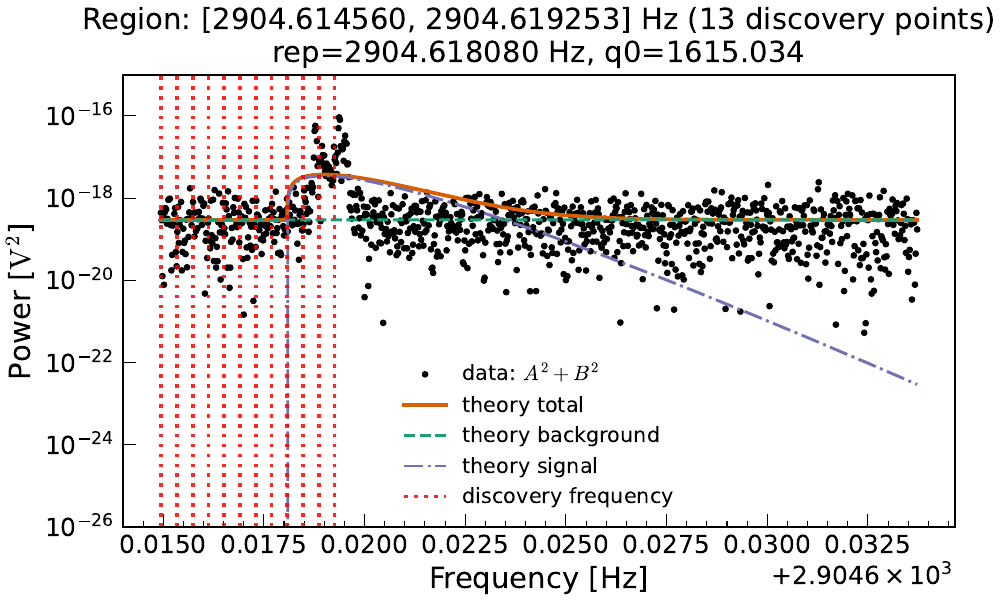}}
\vspace{0.6em}
\makebox[\textwidth][c]{%
\subfloat[]{
\includegraphics[width=0.32\textwidth]{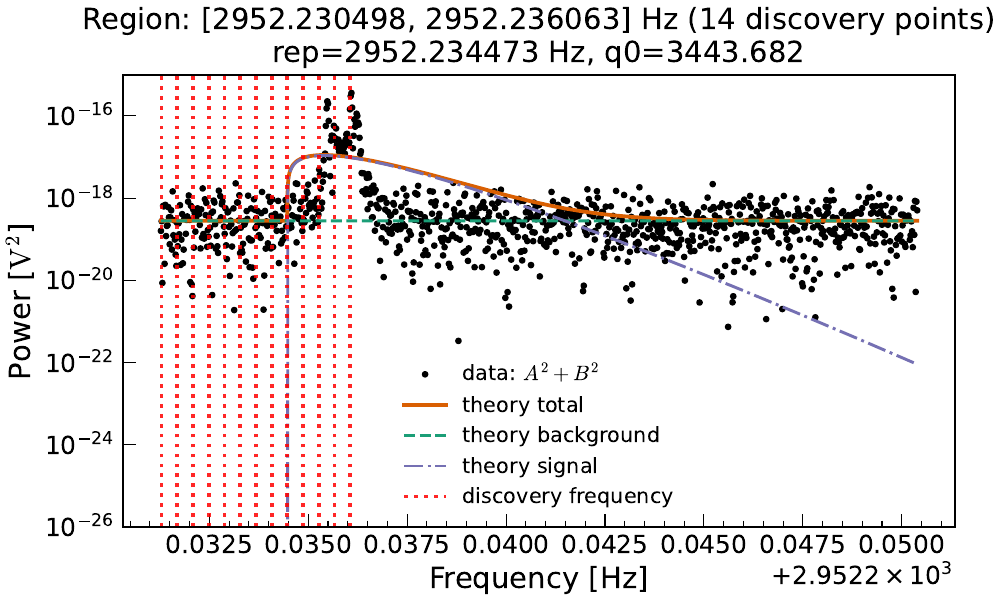}}
\hspace{0.03\textwidth}
\subfloat[]{
\includegraphics[width=0.32\textwidth]{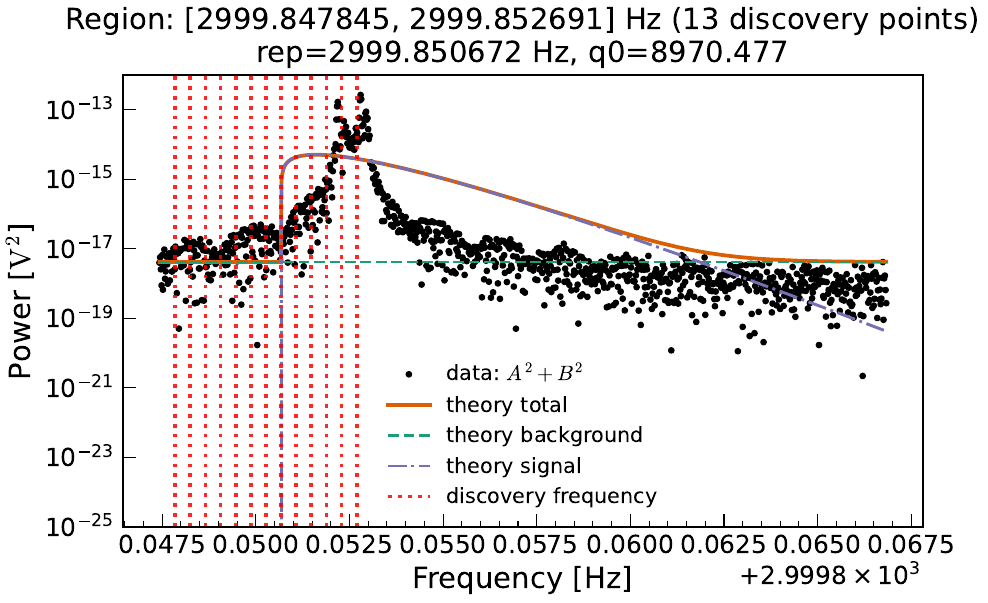}}
}
\caption{
Representative spectra in five high-significance regions.
Black points denote the measured data power $A^2+B^2$.
The red solid curve is the total model spectrum, the green dashed curve is the
instrumental-noise background, and the purple dash-dotted curve is the axion
signal component.
Vertical red lines mark the recurrent candidate frequencies in each region.
The expected axion spectrum is broader than the observed narrow spectral
features, which are instead consistent with instrumental-noise structures.
Therefore, these recurrent high-significance candidates are interpreted as
device-intrinsic noise rather than axion signals.
}
\label{fig:representative_noise_regions}
\end{figure*}

\bibliography{ref-new}
\end{document}